\documentclass[12pt,notitlepage,fleqn]{article}
\usepackage{amssymb}
\usepackage{graphicx}
\usepackage{amsmath}
\usepackage{indentfirst}

\input{tcilatex}

\begin{document}

\begin{center}
\ {\Huge \ Deducing Quark Rest Masses with Phenomenological Formulae }

\bigskip

{\normalsize Jiao Lin Xu}

{\small The Center for Simulational Physics, The Department of Physics and
Astronomy}

{\small University of Georgia, Athens, GA 30602, USA}

E- mail: {\small \ jxu@hal.physast.uga.edu}

\bigskip\ \ \ \ \ \ \ \ \ \ \ \ \ \ \ \ \ \ \ \ \ \ \ \ \ \ \ \ \ \ \ \ \ \
\ \ \ \ \ \ \ \ \ \ \ \ \ \ \ \ 

\ \ \ \ \ \ \ \ \ \ \ \ \ \ \ \ \ \ \ \ \ \ \ \ \ \ \ \ \ \ \ \ \ \ \ \ \ \
\ \ \ \ \ \ \ \ \ \ \ \ \ \ \ \ \ \ \ \ \ 

\textbf{Abstract}
\end{center}

\bigskip

{\small \ Using phenomenological formulae, we can deduce the rest masses and
intrinsic quantum numbers (I, S, C, B and Q) of quarks, baryons and mesons
from only one\ unflavored elementary quark family }$\epsilon ${\small . The
deduced quantum numbers match experimental results exactly, and the deduced
rest masses are 98.5\% (or 97\%) consistent with experimental results for
baryons (or mesons). This paper predicts some quarks \ [d}$_{S}${\small (773)%
}$,$ {\small d}$_{S}${\small (1933) and }$\text{u}_{C}${\small (6073) ],
baryons [}$\Lambda _{c}${\small (6599), }$\Lambda _{b}${\small (9959)]\ and
mesons [D(6231), B(9502)].\ PACS: 12.39.-x; 14.65.-q; 14.20.-c. Keywords:
phenomenological, beyond the standard model.\ \ \ \ }

\section{Introduction}

One hundred years ago, classic physics had already been fully developed.
Most\ physical phenomena could be explained with this physics. Black body
spectrum, however, could not be explained by the physics of that time,
leading Planck to propose a quantization postulate to solve this problem 
\cite{Planck}. The Planck postulate eventually led to quantum mechanics.
Physicists already clearly knew that the black body spectrum was a new
phenomenon outside the applicable area of classic physics. The development
from classic physics to quantum physics depended mainly on new physical
ideas rather than complex mathematics and extra dimensions of space.

Today we face a similar situation. The standard model \cite{Standard}
\textquotedblleft is in excellent accord with almost all current data.... It
has been enormously successful in predicting a wide range of
phenomena,\textquotedblright\ but\ it cannot deduce the mass spectra of
quarks. So far, no theory has been able to successfully do so. Like black
body spectrum, the quark mass spectrum may need a new theory outside the
standard model. M. K. Gaillard, P. D. Grannis, and F. J \ Sciulli have
already pointed out \cite{Standard} that the standard model
\textquotedblleft is incomplete... We do not expect the standard model to be
valid at arbitrarily short distances. However, its remarkable success
strongly suggest that the standard model will remain an \ excellent
approximation to nature at distance scales as small as 10$^{-18}$m... high
degree of arbitrariness suggests that a more fundamental theory underlies
the standard model.\textquotedblright\ The history of quantum physics shows
that a new physics theory's primary need is new physical ideas. This paper
gives new physical ideas using phenomenological formulae. Using these
formulae, we try to deduce the rest masses of quarks.

\section{The Elementary Quarks and Their Free Excited States}

1). We assume that there is only one elementary quark family $\epsilon $
with s = I = $\frac{1}{2}$ and two isospin states ($\epsilon _{u}$ has I$_{Z}
$ = $\frac{1}{2}$ and Q = +$\frac{2}{3}$, $\epsilon _{d}$ has I$_{Z}$ = -$%
\frac{1}{2}$ and Q = -$\frac{1}{3}$). For $\epsilon _{u}$ (or $\epsilon _{d}$%
),\ there are three colored (red, yellow or blue) quarks. Thus, there are
six Fermi elementary quarks in the $\epsilon $ family with S = C = B = 0 in
the vacuum. The elementary quarks $\epsilon _{u}$ and $\epsilon _{d}$ have
the SU(2) symmetries.

2). As a colored elementary quark $\epsilon _{u}$(or $\epsilon _{d}$) is
excited from the vacuum, its color, electric charge, rest mass and spin do
not change, but it will get energy.\ The excited state\ of the elementary
quark $\epsilon _{u}$\ is the u-quark with Q = $\frac{2}{3},$ rest mass m$%
_{u}^{\ast }$, I = s = $\frac{1}{2}$ and I$_{Z}$ = $\frac{1}{2}$. The
excited state\ of the elementary quark $\epsilon _{d}$\ is the d-quark with
Q = - $\frac{1}{3}$, rest mass m$_{d}^{\ast }$, I = s = $\frac{1}{2}$ and I$%
_{Z}$ = $\frac{-1}{2}$. Since $\epsilon _{u}$ and $\epsilon _{d}$ have the
SU(2) symmetries, the free excited quarks u(m$_{u}^{\ast }$) and d(m$%
_{d}^{\ast }$) also have the SU(2) symmetries.

3). According to the Quark Model \cite{Quark Model}, a proton is composed of
three quarks [u(m$_{\text{u}}^{\ast }$)u(m$_{\text{u}}^{\ast }$)d(m$_{\text{d%
}}^{\ast }$)] and a neutron is also composed of three quarks [u(m$_{\text{u}%
}^{\ast }$)u(m$_{\text{d}}^{\ast }$)d(m$_{\text{d}}^{\ast }$)]. Thus proton
mass M$_{\text{p}}$ and neutron mass M$_{\text{n}}$ are:

\begin{equation}
\begin{tabular}{l}
M$_{\text{p}}$ = m$_{\text{u}}$+m$_{\text{u}}$+m$_{\text{d}}-\left\vert 
\text{E}_{\text{Bind}}\text{(p)}\right\vert $, \\ 
M$_{\text{n}}$ = m$_{\text{u}}$+m$_{\text{d}}$+m$_{\text{d}}-\left\vert 
\text{E}_{\text{Bind}}\text{(n)}\right\vert $,%
\end{tabular}
\label{Mp and Mn}
\end{equation}
where $\left\vert \text{E}_{\text{Bind}}\text{(p)}\right\vert $ and $%
\left\vert \text{E}_{\text{Bind}}\text{(n)}\right\vert $ are the strong
binding energy of the three quarks inside p and n. Omitting electromagnetic
masses, we have 
\begin{equation}
\text{M}_{\text{p}}\text{=M}_{\text{n }}\text{= 939 Mev,\ m}_{\text{u}%
}^{\ast }\text{=m}_{\text{d}}^{\ast }\text{ and }\left\vert \text{E}_{\text{%
Bind}}\text{(p)}\right\vert \text{= }\left\vert \text{E}_{\text{Bind}}\text{%
(n)}\right\vert \text{.}  \label{M*m*}
\end{equation}
$\left\vert \text{E}_{\text{Bind}}\text{(p)}\right\vert $= $\left\vert \text{%
E}_{\text{Bind}}\text{(n)}\right\vert $= $\left\vert \text{E}_{\text{Bind}%
}\right\vert $ is a unknown complex function. As a phenomenological
approximation, we assume that $\left\vert \text{E}_{\text{Bind}}\right\vert $
= 3$\Delta $. $\Delta $\ is an unknown large constant ($\Delta $ \TEXTsymbol{%
>}\TEXTsymbol{>} m$_{\text{P}}$= 938 Mev). From (\ref{Mp and Mn}) and (\ref%
{M*m*}), we find m$_{\text{u}}^{\ast }$= m$_{\text{d}}^{\ast }$ = 313 + $%
\Delta $

\begin{equation}
\begin{tabular}{l}
$\text{u(313+}\Delta \text{) \ \ and \ \ d(313+}\Delta \text{) }$ \\ 
$\text{\ }\Delta $ = $\frac{1}{3}\left\vert \text{E}_{\text{Bind}%
}\right\vert $ \TEXTsymbol{>}\TEXTsymbol{>}m$_{\text{P}}$=938Mev%
\end{tabular}
\label{313&Dalta}
\end{equation}
\qquad \qquad

Remember that we have already found two long-lived quarks u(313+$\Delta $)
and d(313+$\Delta $). They are the free excited states of $\epsilon _{u}$
and $\epsilon _{d}$. They compose the most important baryons p(939) (with I
= $\frac{1}{2}$. I$_{z}$= $\frac{1}{2}$, Q =1, S = C = B = 0) and n(939)
(with I = $\frac{1}{2}$. I$_{z}$= $\frac{-1}{2}$, Q = S = C = B = 0) and
mesons $\pi $ (with I =1, Q =+1, 0, -1, S = C = B = 0).

\section{Phenomenological Formulae\ for Energy Bands\ \ \ \ \ \ \ \ \ \ \ \
\ \ \ \ \ \ \ \ \ \ \ \ \ \ \ \ \ \ \ \ \ \ \ \ \ \ \ \ \ \ \ \ \ \ \ \ \ \
\ \ \ \ \ \ \ \ \ \ \ \ \ \ \ \ \ \ \ \ \ \ \ \ \ \ \ \ \ \ \ \ \ \ \ \ \ \
\ \ \ \ \qquad \qquad \qquad}

1). In order to deduce the short-lived quarks, we assume a phenomenological
energy band formula. There are energy band excited states of the elementary
quark $\epsilon $ whose energies are given by the following formula: 
\begin{equation}
\begin{tabular}{l}
$\mathbb{E}\text{(}\overrightarrow{\kappa }\text{,}\overrightarrow{n}\text{)
= 313 + }\Delta \text{ + 360 E(}\overrightarrow{\kappa }\text{,}%
\overrightarrow{n}\text{)}$ \\ 
E($\overrightarrow{\kappa }$,$\overrightarrow{n}$) = [(n$_{1}$-$\xi $)$^{2}$%
+(n$_{2}$-$\eta $)$^{2}$+(n$_{3}$-$\zeta $)$^{2}$],%
\end{tabular}
\label{E(nk)}
\end{equation}%
\ where $\overrightarrow{\kappa }$ = ($\xi $, $\eta $, $\zeta $). ($\xi $, $%
\eta $, $\zeta $) are the coordinates of the symmetry axes of the regular
rhombic dodecahedron in $\overrightarrow{\kappa }$-space as shown in Fig.1
and $\overrightarrow{n}$ = (n$_{1}$, n$_{2}$, n$_{3}$), n$_{1}$, n$_{2}$ and
n$_{3}$ are $\pm \func{integer}$s or zero.\ \ \ \ \ \ \ \ \ \ \ \ \ \ \ \ \
\ \ \ \ \ \ \ \ \ 

\ \ \ \ \ \ \ \ \ \ \ \ \ \ \ 

\begin{figure}[h]
\vspace{5.8in} \includegraphics{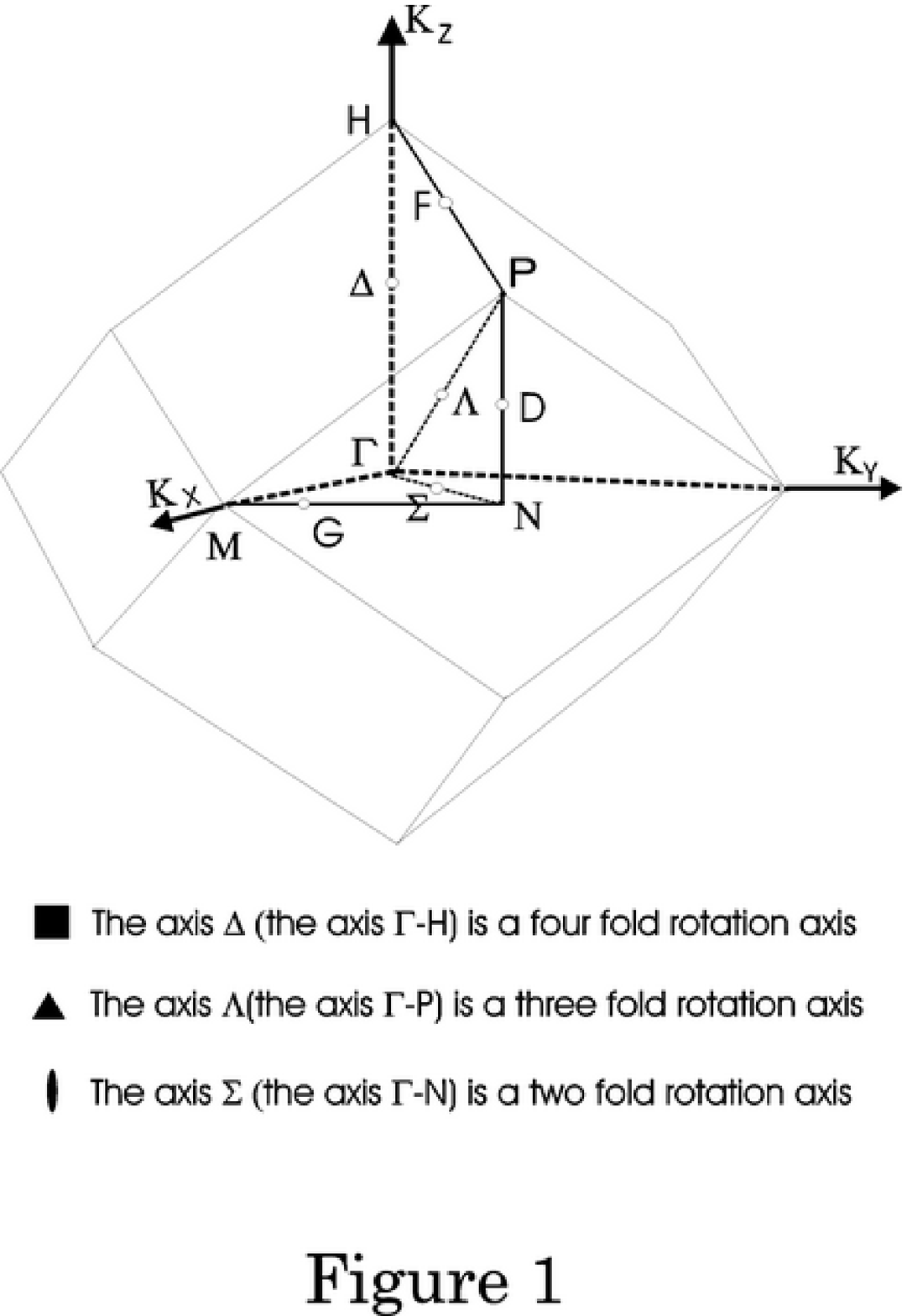} \label{Fig1}
\caption{{\protect\small The regular rhombic dodecahedron. The symmetry
points and axes are indicated.}}
\end{figure}

\ \ \ \ \ \ \ \qquad \qquad

2). If we assume n$_{1}$ = \textit{l}$_{2}$ \textit{+ l}$_{3}$, n$_{2}$ =%
\textit{\ l}$_{3}$ \textit{+ l}$_{1}$ and n$_{3}$ =\textit{\ l}$_{1}$ 
\textit{+ l}$_{2},$ so that 
\begin{equation}
\begin{tabular}{l}
\textit{l}$_{1}$ = $\frac{1}{2}$(-n$_{1}$ + n$_{2}$ + n$_{3}$) \\ 
\textit{l}$_{2}$ = $\frac{1}{2}$(+n$_{1}$ - n$_{2}$ + n$_{3}$) \\ 
\textit{l}$_{3}$ = $\frac{1}{2}$(+n$_{1}$ + n$_{2}$ - n$_{3}$).%
\end{tabular}
\label{l-n}
\end{equation}%
n$_{1}$, n$_{2}$ and n$_{3}$ are those values of $\overrightarrow{n}$ = (n$%
_{1}$, n$_{2}$, n$_{3}$) that make $\overrightarrow{l}$ = \textit{(l}$_{1}$%
\textit{, l}$_{2}$\textit{, l}$_{3}$\ ) an integer vector (\textit{l}$_{1}$%
\textit{, l}$_{2}$\textit{, l}$_{3}$ are $\pm \func{integer}$s or zero). For
example, $\vec{n}$ \ cannot take the values (1, 0, 0) or (1, 1, -1), but can
take (0, 0, 2) and (1, -1, 2). From E($\overrightarrow{\kappa }$,$%
\overrightarrow{n}$), we can give a definition of the equivalent $%
\overrightarrow{n}$: for $\xi $ = $\eta $ = $\varsigma $ = 0, all $%
\overrightarrow{n}$ values that give the same E($\overrightarrow{\kappa }$,$%
\overrightarrow{n}$) value are equivalent n-values. We show the low level
equivalent $\overrightarrow{n}$-values that satisfy condition (\ref{l-n}) in
the following list (note $\overline{\text{n}_{i}}$ = - n$_{i}$):\ 

\begin{equation}
\begin{tabular}{|l|}
\hline
{\small E(0, }$\overrightarrow{n}${\small ) = 0\ : (0, 0, 0) \ \ \ \ \ \ \ \
\ \ Notes: }[$\overline{\text{{\small 1}}}${\small 12 }$\equiv $ (-1,1,2)
and $\overline{\text{{\small 1}}}${\small 1}$\overline{\text{{\small 2}}}%
\equiv $ (-1,1,-2)] \\ \hline
{\small E(0, }$\overrightarrow{n}${\small ) = 2\ :}$\ ${\small (101, }$%
\overline{\text{{\small 1}}}${\small 01, 011, 0}$\overline{\text{{\small 1}}}
${\small 1, 110, 1}$\overline{\text{{\small 1}}}${\small 0, }$\overline{%
\text{{\small 1}}}${\small 10, }$\overline{\text{{\small 1}}}\overline{\text{%
{\small 1}}}${\small 0, 10}$\overline{\text{{\small 1}}}${\small , }$%
\overline{\text{{\small 1}}}${\small 0}$\overline{\text{{\small 1}}}${\small %
, 01}$\overline{\text{{\small 1}}}${\small , 0}$\overline{\text{{\small 1}}}%
\overline{\text{{\small 1}}}${\small )} \\ \hline
{\small E(0, }$\overrightarrow{n}${\small ) = 4\ :\ \ (002, 200}, {\small 200%
}$\text{, }\overline{\text{{\small 2}}}\text{{\small 00}, }${\small 0}$%
\overline{\text{{\small 2}}}${\small 0, 00}$\overline{\text{{\small 2}}}$%
{\small )} \\ \hline
{\small E(0, }$\overrightarrow{n}${\small ) = 6:} 
\begin{tabular}{l}
{\small 112, 211, 121, }$\overline{\text{{\small 1}}}${\small 21,}$\overline{%
\text{{\small 1}}}${\small 12, 2}$\overline{\text{{\small 1}}}${\small 1}, 
{\small 1}$\overline{\text{{\small 1}}}${\small 2, 21}$\overline{\text{%
{\small 1}}}${\small ,12}$\overline{\text{{\small 1}}}${\small ,}$\overline{%
\text{{\small 2}}}${\small 11, 1}$\overline{\text{{\small 2}}}${\small 1, 11}%
$\overline{\text{{\small 2}}}$,{\small \ } \\ 
$\overline{\text{{\small 11}}}${\small 2, }$\overline{\text{{\small 1}}}$%
{\small 2}$\overline{\text{{\small 1}}}$,{\small \ 2}$\overline{\text{%
{\small 11}}}${\small , }$\overline{\text{{\small 21}}}${\small 1}, $%
\overline{\text{{\small 12}}}${\small 1, 1}$\overline{\text{{\small 12}}}$%
{\small , 1}$\overline{\text{{\small 21}}}${\small ,}$\overline{\text{%
{\small 1}}}${\small 1}$\overline{\text{{\small 2}}}${\small ,}$\overline{%
\text{{\small 2}}}$1$\overline{\text{{\small 1}}}$,{\small \ }$\overline{%
\text{{\small 211}}}${\small , }$\overline{\text{{\small 121}}},${\small \ }$%
\overline{\text{{\small 112}}},$%
\end{tabular}
\\ \hline
{\small E(0, }$\overrightarrow{n}${\small ) = 8:}$\ ${\small (220, 2}$%
\overline{\text{{\small 2}}}${\small 0, }$\overline{\text{{\small 2}}}$%
{\small 20, }$\overline{\text{{\small 2}}}\overline{\text{{\small 2}}}$%
{\small 0, 202, 20}$\overline{\text{{\small 2}}}${\small , }$\overline{\text{%
{\small 2}}}${\small 02, }$\overline{\text{{\small 2}}}${\small 0}$\overline{%
\text{{\small 2}}}${\small , 022, 02}$\overline{\text{{\small 2}}}${\small ,
0}$\overline{\text{{\small 2}}}${\small 2, 0}$\overline{\text{{\small 2}}}%
\overline{\text{{\small 2}}}${\small )} \\ \hline
\end{tabular}
\label{nnn}
\end{equation}
$\ $\ \ \ \ \ \ \ \ \ \ \ \ \ 

3). From Fig.1, we can see that there are four kinds of symmetry points ($%
\Gamma $, H, P and N) and six kinds of symmetry axes ($\Delta $, $\Lambda $, 
$\Sigma $, D, F and G) in the regular rhombic dodecahedron. The coordinates (%
$\xi $, $\eta $, $\varsigma $) of the symmetry points are: 
\begin{equation}
\overrightarrow{\kappa }_{_{\Gamma }}\text{=}\text{ (0, 0, 0), }%
\overrightarrow{\kappa }_{\text{H}}\text{ = (0, 0, 1), }\overrightarrow{%
\kappa }_{\text{ P}}\text{=}\text{(}\frac{\text{1}}{\text{2}}\text{, }\frac{%
\text{1}}{\text{2}}\text{, }\frac{\text{1}}{\text{2}}\text{), }%
\overrightarrow{\kappa }_{\text{N}}\text{ }\text{= (}\frac{\text{1}}{\text{2}%
}\text{, }\frac{\text{1}}{\text{2}}\text{, 0). }  \label{S-Point}
\end{equation}%
The coordinates ($\xi $, $\eta $, $\varsigma $) of the symmetry axes are: 
\begin{equation}
\begin{tabular}{ll}
$\overrightarrow{\kappa }_{\Delta }\text{ }\text{= (0, 0, }\zeta \text{),\ \
0}\leq \zeta \text{ }\leq \text{1; }$ & $\overrightarrow{\kappa }_{\Lambda }%
\text{ = (}\xi \text{, }\xi \text{, }\xi \text{), \ 0}\leq \xi \leq \frac{%
\text{1}}{\text{2}}\text{;}$ \\ 
$\overrightarrow{\kappa }_{\Sigma }\text{{} }\text{= (}\xi \text{, }\xi 
\text{, 0), \ 0}\leq \xi \text{ }\leq \frac{\text{1}}{\text{2}}\text{;}$ & $%
\overrightarrow{\kappa }_{\text{D}}\text{ }\text{= (}\frac{\text{1}}{\text{2}%
}\text{, }\frac{\text{1}}{\text{2}}\text{, }\xi \text{), \ 0}\leq \xi \leq 
\frac{\text{1}}{\text{2}}\text{;}$ \\ 
$\overrightarrow{\kappa }_{\text{G}}\text{ }\text{= (}\xi \text{, 1-}\xi 
\text{, 0), \ }\frac{\text{1}}{\text{2}}\leq \xi \text{ }\leq \text{1;}$ & $%
\overrightarrow{\kappa }_{\text{F}}\text{ = (}\xi \text{, }\xi \text{, 1-}%
\xi \text{), \ 0}\leq \xi \leq \frac{\text{1}}{\text{2}}\text{.}$%
\end{tabular}
\label{Sym-Axes}
\end{equation}%
\qquad 

4). The energy (\ref{E(nk)}) with an allowed $\overrightarrow{n}$\ = (n$_{1}$%
, n$_{2}$, n$_{3}$) in (\ref{nnn}) along a symmetry axis [$\overrightarrow{%
\kappa }$ = ($\xi $,$\eta $,$\varsigma $) take the values in (\ref{Sym-Axes}%
)] forms an energy band. Each energy band corresponds to a short-lived quark.

5). After getting (\ref{E(nk)}), (\ref{nnn}), (\ref{S-Point}) and (\ref%
{Sym-Axes}), we can deduce low energy bands of the six symmetry axes (see
Appendix B of \cite{0502091})). As an example, we will deduce the single
energy bands of the $\Delta $-axis in this paper.\ For the $\Delta $-axis, $%
\overrightarrow{\kappa }_{\Delta }$ = (0, 0, $\zeta $) from (\ref{Sym-Axes}%
). Putting $\overrightarrow{\kappa }_{\Delta }$\ into (\ref{E(nk)}), we get $%
\mathbb{E}$($\overrightarrow{\kappa }$,$\overrightarrow{n}$)$_{\Delta }$
=313 + $\Delta $ + 360[(n$_{1}$)$^{2}$+(n$_{2}$)$^{2}$+(n$_{3}$-$\zeta $)$%
^{2}$]. For point-$\Gamma ,\overrightarrow{\kappa }_{\Gamma }$\ = (0, 0, $0$%
) from (\ref{S-Point}), E$_{\Gamma }$($\overrightarrow{n}$) = (n$_{1}$)$^{2}$%
+(n$_{2}$)$^{2}$+(n$_{3}$)$^{2}$. For point-H$,\overrightarrow{\kappa }_{%
\text{H}}$\ = (0, 0, 1) from (\ref{S-Point}), E$_{\text{H}}$($%
\overrightarrow{n}$) = (n$_{1}$)$^{2}$+(n$_{2}$)$^{2}$+(n$_{3}$-1)$^{2}$.
Putting $\text{(n}_{1}\text{, n}_{2}\text{, n}_{3}${\small ) }values of the
single bands of the $\Delta $-axis into $\mathbb{E}$($\overrightarrow{\kappa 
}$,$\overrightarrow{n}$)$_{\Delta }$, E$_{\Gamma }$($\overrightarrow{n}$)
and E$_{\text{H}}$($\overrightarrow{n}$), we can find energy bands as shown
in Table 1:

\ \ \ \ \ \ \ \ \ \ \ \ \ \ \ \ \ \ 

\begin{tabular}{|l|}
\hline
\ \ \ \ Table 1\ The Single Energy Bands of the $\Delta $-Axis (the $\Gamma $%
-H axis)\  \\ \hline
\ $\mathbb{E}$($\overrightarrow{\kappa }$,$\overrightarrow{n}$)$_{\Delta }$%
{\small \ =313+}$\Delta ${\small +360[(n}$_{1}${\small )}$^{2}${\small +(n}$%
_{2}${\small )}$^{2}${\small +(n}$_{3}${\small -}$\zeta ${\small )}$^{2}$%
{\small ] }\ $[\text{\ 0 }\leq \text{ }\zeta \text{ }\leq \text{1}]$ \\ 
\hline
E$_{\Gamma }$($\overrightarrow{n}$) = (n$_{1}$)$^{2}$+(n$_{2}$)$^{2}$+(n$%
_{3} $)$^{2}$ and E$_{\text{H}}$($\overrightarrow{n}$) = (n$_{1}$)$^{2}$+(n$%
_{2}$)$^{2}$+(n$_{3}$-1)$^{2}$ \\ \hline
$%
\begin{tabular}{|l|l|l|l|l|}
\hline
(n$_{1}$,n$_{2}$,n$_{3}${\small )} & E($\overrightarrow{\kappa }$,$%
\overrightarrow{n}$)$_{Start}$ & Minim.E & {\small E-Band\ }$\mathbb{E}$($%
\overrightarrow{\kappa }$,$\overrightarrow{n}$)$_{\Delta }${\small \ } & E($%
\overrightarrow{\kappa }$,$\overrightarrow{n}$)$_{end}$ \\ \hline
{\small (0, 0, 0)} & E$_{\Gamma }${\small (0,0,0)}= 0 & {\small 313+}$\Delta 
$ & {\small 313+}$\Delta ${\small +}$\zeta ^{2}$ & E$_{H}${\small (0,0,0)}=1
\\ \hline
{\small (0, 0, 2)} & E$_{H}${\small (0,0,2)}= 1 & {\small 673}+$\Delta $ & 
313+$\Delta $+(2-$\zeta $)$^{2}$ & E$_{\Gamma }${\small (0,0,2)}=4 \\ \hline
{\small (0, 0, -2)} & E$_{\Gamma }${\small (0,0,-}2{\small )}=4 & {\small %
1753}+$\Delta $ & 313+$\Delta $+(2+$\zeta $)$^{2}$ & E$_{H}${\small (0,0,-}2%
{\small )}=9 \\ \hline
{\small (0, 0, 4)} & E$_{H}${\small (0,0,4)}= 9 & {\small 3553}+$\Delta $ & 
313+$\Delta $+(4-$\zeta $)$^{2}$ & E$_{\Gamma }${\small (0,0,4)}=16 \\ \hline
{\small (0, 0, -4)} & E$_{\Gamma }${\small (0,0,-4)}=16 & {\small 6073}+$%
\Delta $ & 313+$\Delta $+(4+$\zeta $)$^{2}$ & E$_{H}${\small (0,0,-4)}=25 \\ 
\hline
{\small (0, 0, 6)} & E$_{H}${\small (0,0,6)}=25 & {\small 9313}+$\Delta $ & 
313+$\Delta $+(6-$\zeta $)$^{2}$ & E$_{\Gamma }${\small (0,0,6)}=36 \\ \hline
\end{tabular}%
\ \ \ $ \\ \hline
$\text{E(}\overrightarrow{\kappa }\text{,}\overrightarrow{n}\text{)}_{Start}$%
{\small is the value of }$\text{E(}\overrightarrow{\kappa }\text{,}%
\overrightarrow{n}\text{) at the start point of the energy band}$ \\ \hline
$\text{E(}\overrightarrow{\kappa }\text{,}\overrightarrow{n}\text{)}_{end}$%
{\small is the value of }$\text{E(}\overrightarrow{\kappa }\text{,}%
\overrightarrow{n}\text{) at the end point of the energy band}$ \\ \hline
\end{tabular}

\ \ \ \ \ \ \ \ \ \ \ \ \ \ \ \ \ \ \ \ \ \ \ \ \ \ \ \ \ \ \ \ \ \ \ \ 

Similarly, we can deduce the single energy bands of the $\Sigma $-axis. For
the $\Sigma $-axis, $\overrightarrow{\kappa }_{\Sigma }$= ($\xi $, $\xi $, $%
0 $). Putting the$\overrightarrow{\kappa }_{\Sigma }$ into (\ref{E(nk)}), we
have {\small \ }$\mathbb{E}$($\overrightarrow{\kappa }$,$\overrightarrow{n}$)%
$_{\Sigma }$ = 313+ $\Delta $ + $360$[(n$_{1}$-$\xi $)$^{2}$+(n$_{2}$-$\xi $)%
$^{2}$+(n$_{3}$)$^{2}$] $\text{\ 0}\leq \zeta \leq \frac{\text{1}}{2}$. For
point-N$,$ $\overrightarrow{\kappa }_{\text{N}}$\ = ($\frac{1}{2}$, $\frac{1%
}{2}$, 0) from (\ref{S-Point}), E$_{\text{N}}$($\overrightarrow{n}$) = (n$%
_{1}$-$\frac{1}{2}$)$^{2}$+(n$_{2}$-$\frac{1}{2}$)$^{2}$+(n$_{3}$)$^{2}$.
Putting $\text{(n}_{1}\text{,n}_{2}\text{,n}_{3}${\small ) }values of the
single bands of the $\Sigma $-axis (Appendix B, Table B2 of \cite{0502091})
into E($\vec{k}$,$\vec{n}$), {\small E}$_{\Gamma }${\small (}$%
\overrightarrow{n}${\small ) and E}$_{\text{N}}${\small (}$\overrightarrow{n}
${\small ), }we can deduce energy bands as shown in Table 2.

\ \ \ \ \ \ \ \ \ \ \ \ \ \ \ \ \ \ \ \ \ \ 

\begin{tabular}{|l|}
\hline
\ \ \ \ Table 2\ The Single Energy Bands of the $\Sigma $-Axis (the $\Gamma $%
-N axis)\ $\text{ }$ \\ \hline
{\small Energy Band\ \ }$\mathbb{E}$($\overrightarrow{\kappa }$,$%
\overrightarrow{n}$)$_{\Sigma }${\small \ =313+ }$\Delta ${\small \ + }$360$%
{\small [(n}$_{1}${\small -}$\xi ${\small )}$^{2}${\small +(n}$_{2}${\small -%
}$\xi ${\small )}$^{2}${\small +(n}$_{3}${\small )}$^{2}${\small ] }$_{\text{%
\ (0}\leq \zeta \leq \frac{\text{1}}{2})}$ \\ \hline
\ $%
\begin{tabular}{|l|l|l|l|l|}
\hline
\ n$_{1}$n$_{2}$n$_{3}$ & E($\overrightarrow{\kappa }$,$\overrightarrow{n}$)$%
_{Start}$ & minim.E & {\small \ E-Band}$\mathbb{E}$($\overrightarrow{\kappa }
$,$\overrightarrow{n}$)$_{\Sigma }$ & E($\overrightarrow{\kappa }$,$%
\overrightarrow{n}$)$_{end}$ \\ \hline
{\small (0, 0, 0)} & E$_{\Gamma }${\small (0, 0, 0)}=0 & {\small 313}+$%
\Delta $ & {\small 313+}$\Delta ${\small +720}$\xi ^{2}$ & E$_{N}${\small %
(0, 0, 0)}=$\frac{\text{1}}{2}$ \\ \hline
({\small 1, 1, 0}) & E$_{N}$({\small 1, 1, 0})=$\frac{\text{1}}{2}$ & 
{\small 493+}$\Delta $ & {\small 313+}$\Delta ${\small +720(1-}$\xi ${\small %
)}$^{2}$ & E$_{\Gamma }$({\small 1, 1, 0})=2 \\ \hline
(-{\small 1,-1,0}) & E$_{\Gamma }$(-{\small 1,-1,0})=2 & {\small 1033+}$%
\Delta $ & {\small 313+}$\Delta ${\small +720(1+}$\xi ${\small )}$^{2}$ & E$%
_{N}$(-{\small 1,-1,0})=$\frac{\text{9}}{2}$ \\ \hline
({\small 2, 2, 0}) & E$_{N}$({\small 2, 2, 0})=$\frac{\text{9}}{2}$ & 
{\small 1933+}$\Delta $ & {\small 313+}$\Delta ${\small +720(2-}$\xi $%
{\small )}$^{2}$ & E$_{\Gamma }$({\small 2, 2, 0})=8 \\ \hline
(-{\small 2,-2,0}) & E$_{\Gamma }$(-{\small 2,-2,0})=8 & {\small 3193+}$%
\Delta $ & {\small 313+}$\Delta ${\small +720(2+}$\xi ${\small )}$^{2}$ & E$%
_{N}$(-{\small 2,-2,0})=$\frac{\text{25}}{2}$ \\ \hline
{\small (3, 3, 0)} & $\text{E}_{N}${\small (3, 3, 0)}$\text{=}\frac{\text{25}%
}{2}$ & $\text{4813}${\small +}$\Delta $ & {\small 313+}$\Delta ${\small %
+720(3-}$\xi ${\small )}$^{2}$ & $\text{E}_{\Gamma }${\small (3, 3, 0)}$%
\text{=18}$ \\ \hline
{\small (-3,-3,0)} & $\text{E}_{\Gamma }${\small (-3,-3,0)}$\text{=18}$ & $%
\text{6793}${\small +}$\Delta $ & {\small 313+}$\Delta ${\small +720(30+}$%
\xi ${\small )}$^{2}$ & $\text{E}_{N}${\small (-3,-3,0)}$\text{=}\frac{\text{%
49}}{2}$ \\ \hline
{\small (4, 4, 0)} & $\text{E}_{N}${\small (4, 4, 0)}$\text{=}\frac{\text{49}%
}{2}$ & $\text{9133}${\small +}$\Delta $ & {\small 313+}$\Delta ${\small %
+720(3-}$\xi ${\small )}$^{2}$ & $\text{E}_{\Gamma }${\small (4, 4, 0)}$%
\text{=32}$ \\ \hline
\end{tabular}%
\ \ \ $ \\ \hline
\end{tabular}%
\ \ \ \ \ \ \ \ \ \ \ \ \ \ \ \ \ \ \ \ \ \ \ \ \ \ \ \ \ \ \ 

\section{Phenomenological Formulae\ for Rest Masses and Intrinsic Quantum
Numbers \ \ \ \ \ \ \ \ \ \ \ \ \ \ \ \ \ \ \ \ \ \ \ \ \ \ \ \ \ \ \ \ \ \
\ \ \ \ \ \ \ \ \ \ \ \ \ \ \ \ \ \ \ \ \ \ \ \ \ \ \ \ \ \ \ \ \ \ \ \ \ \
\ \ \ \ \ \ \ \ \ \ \ \ \ \ \ \ \ \ \ \ \ \ \ \ \ \ \ \ \ \ \ \ \ \ \ \ \ \
\ \ \ \ \ \ \ \ \ \ \ \ \ \ \ \ \ \ \ \ \ \ \ \ \ \ \ \ \ \ \ \ \ \ \ \ \ \
\ \ \ \ \ \ \ \ \ \ \ \ \ \ \ \ \ \ \ \ \ \ \ \ \ \ \ \ \ \ \ \ \ \ \ \ \ \
\ \ \ \ \ \ \ \ \ \ \ \ \ \ \ \ \ \ \ \ \ \ \ \ \ \ \ \ \ \ \ \ \ \ \ \ \ \
\ \ \ \ \ \ \ \ \ \ \ \ \ \ \ \ \ \ \ \ \ \ \ \ \ \ \qquad\ \ \ \ \ \ \ \ \
\ \ \ \ \ \ \ \ \ \ \ \ \ \ \ \ \ \ \ \ \ \ \ \ \ \ \ \ \ \ \ \ \ \ \ \ \ \
\ \ \ \ \ \ \ \ \ \ \ \ \ \ \ \ \ \ \ \ \ \ \ \ \ \ \ \ \ \ \ \ \ \ \ \ \ \
\ \ \ \ \ \ \ \ \ \ \ \ \ \ \ \ \ \ \ \ \ \ \ \ \ \ \ \ \ \ \ \ \ \ \ \ \ \
\ \ \ \ \ \ \ \ \ \ \ \ \ \ \ \ \ \ \ \ \ \ \ \ \ \ \ \ \ \ \ \ \ \ \ \ \ \
\ \ \ \ \ \ \ \ \ \ \ \ \ \ \ \ \ \ \ \ \ \ \ \ \ \ \ \ \ \ \ \ \ \ \ \ \ \
\ \ \ \ \ \ \ \ \ \ \ \ \ \ \ \ \ \ \ \ \ \ \ \ \ \ \ \ \ \ \ \ \ \ \ \ \ \
\ \ \ \ \ \ \ \ \ \ \ \ \ \ \ \ \ \ \ \ \ \ \ \ \ \ \ \ \ \ \ \ \ \ \ \ \ \
\ \ \ \ \ \ \ \ \ \ \ \ \ \ \ \ \ \ \ \ \ \ \ \ \ \ \ \ \ \ \ \ \qquad\ \ \
\ \ \ \ \ \ \ \ \ \ \ \ \ \ \ \ \ \ \ \ \ \ \ \ \ \ \ \ \ \ \ \ \ \ \ \ \ \
\ \ \ \ \ \ \ \ \ \ \ \ \ \ \ \ \ \ \ \ \ \ \ \ \ \ \ \ \ \ \ \ \ \ \ \ \ \
\ \ \ \ \ \ \ \ \ \ \ \ \ \ \ \ \ \ \ \ \ \ \ \ \ \ \ \ \ \ \ \ \ \ \ \ \ \
\ \ \ \ \ \ \ \ \ \ \ \ \ \ \ \ \ \ \ \ \ \ \ \ \ \ \ \ \ \ \ \ \ \ \ \ \ \
\ \ \ \ \ \ \ \ \ \ \ \ \ \ \ \ \ \ \ \ \ \ \ \ \ \ \ \ \ \ \ \ \ \ \ \ \ \
\ \ \ \ \ \ \ \ \ \ \ \ \ \ \ \ \ \ \ \ \ \ \ \ \ \ \ \ \ \ \ \ \ \ \ \ \ \
\ \ \ \ \ \ \ \ \ \ \ \ \ \ \ \ \ \ \ \ \ \ \ \ \ \ \ \ \ \ \ \ \ \ \ \ \ \
\ \ \ \ \ \ \ \ \ \ \ \ \ \ \ \ \ \ \ \ \ \ \ \ \ \ \ \ \ \ \ \ \ \ \ \ \ \
\ \ \ \ \ \ \ \ \ \ \ \ \ \ \ \ \ \ \ \ \ \ \ \ \ \ \ \ \ \ \ \ \ \ \ \ \ \
\ \ \ \ \ \ \ \ \qquad\ \ \ \ \ \ \ \ \ \ \ \ \ \ \ \ \ \ \ \ \ \ \ \ \ \ \
\ \ \ \ \ \ \ \ \ \ \ \ \ \ \ \ \ \ \ \ \ \ \ \ \ \ \ \ \ \ \ \ \ \ \ \ \ \
\ \qquad}

In order to deduce the\textbf{\ }short-lived quarks from the energy bands in
Tables 1 and 2, we assume that each energy band corresponds to a quark and
that the rest mass and intrinsic quantum numbers (I, S, C, B and Q) of the
quarks can be deduced using the following phenomenological formulae from the
energy bands:

1). For a group of degenerate energy bands (number = deg) with the same
energy and equivalent $\overrightarrow{n}$ values (\ref{nnn}), the isospin
of the corresponding excited quark is

\begin{equation}
2\text{I+1 = deg}\rightarrow \text{I = }\frac{\text{deg - 1}}{\text{2}}
\label{IsoSpin}
\end{equation}

2). The strange number S of an excited quark that lies on an axis with a
rotary fold R of the regular rhombic dodecahedron is 
\begin{equation}
\text{S = R - 4.}  \label{S-Number}
\end{equation}

3). For an energy band with deg \TEXTsymbol{<} R and R - deg $\neq $ 2, the
strange number of the corresponding quark is

\begin{equation}
\text{%
\begin{tabular}{l}
$\text{deg\TEXTsymbol{<}R and R-deg}\neq \text{2, \ S = S}_{axis}\text{+}%
\Delta \text{S,}$ \\ 
$\Delta \text{S = }\delta \text{(}\widetilde{n}\text{) + [1-2}\delta \text{(S%
}_{axis}\text{)]Sign(}\widetilde{n}\text{)}$%
\end{tabular}
\ }  \label{S+DS}
\end{equation}
where $\delta $($\widetilde{n}$)\ and $\delta $(S$_{axis}$) are Dirac
functions and S$_{axis}$ is the strange number (\ref{S-Number}) of the axis.
For an energy band with $\overrightarrow{n}$ = (n$_{1}$, n$_{2}$, n$_{3})$, $%
\widetilde{n}$ \ is defined as 
\begin{equation}
\widetilde{n}\text{ }\equiv \frac{\text{n}_{1}\text{+n}_{2}\text{+n}_{3}}{%
\left\vert \text{n}_{1}\right\vert \text{+}\left\vert \text{n}%
_{2}\right\vert \text{+}\left\vert \text{n}_{3}\right\vert }\text{, Sgn(}%
\widetilde{n}\text{) = }\left[ \text{%
\begin{tabular}{l}
+1 for $\widetilde{n}$ \TEXTsymbol{>} 0 \\ 
0 \ \ for $\widetilde{n}$ = 0 \\ 
-1 \ for $\widetilde{n}$ \TEXTsymbol{<} 0%
\end{tabular}
}\right] \text{.}  \label{n/n}
\end{equation}

\begin{equation}
\text{If }\widetilde{n}\text{ = 0, \ \ \ \ }\Delta \text{S = }\delta \text{%
(0) = +1 \ from (\ref{S+DS}) and (\ref{n/n}).}  \label{n=0-DS=+1}
\end{equation}

\begin{equation}
\text{If \ }\widetilde{n}\text{ = }\frac{0}{0}\text{, }\Delta \text{S = - S}%
_{Axis}\text{ .}  \label{DaltaS=-Sax}
\end{equation}
Thus, for $\overrightarrow{n}$ = (0, 0, 0), from (\ref{DaltaS=-Sax}), we
have \ 
\begin{equation}
\text{S = S}_{Axis}\text{+}\Delta \text{S = S}_{Axis}\text{- S}_{Axis}\text{
= 0.}  \label{S=0 of n=0}
\end{equation}
\qquad \qquad \qquad\ \ \ \ 

4). If S = +1, we call it the charmed number C (= 1): 
\begin{equation}
\text{if }\Delta \text{S = +1}\rightarrow \text{S =S}_{Ax}\text{+}\Delta 
\text{S = +1, }C\text{ }\equiv \text{+1.}  \label{Charmed}
\end{equation}%
\qquad \qquad

\ \ \ \ If S = -1$,$ which originates from $\Delta S=+1$ on a single energy
band (S$_{Ax}$= -2), and there is an energy fluctuation (\ref{Dalta-E}),$\ $%
we call it the bottom number\ B:$\ \ \ \ \ \ \ \ \ \ \ \ \ $%
\begin{equation}
\text{for single bands, if\ }\Delta \text{S = +1}\rightarrow \text{S = -1
and }\Delta \text{E}\neq \text{0, B}\equiv \text{-1.}  \label{Battom}
\end{equation}

5). The elementary quark $\epsilon _{u}$ (or $\epsilon _{d}$) determines the
electric charge Q of an excited quark. For an excited quark of $\epsilon
_{u} $ (or $\epsilon _{d}$), Q = +$\frac{2}{3}$ (or -$\frac{1}{3}$). For an
excited quark with isospin I, there are 2I +1 members . I$_{z}$ \TEXTsymbol{>%
} 0, Q = +$\frac{2}{3}$; I$_{z}$ \TEXTsymbol{<} 0, Q = -$\frac{1}{3}$; 
\begin{eqnarray}
\text{for I}_{z} &\text{=}&\text{ 0, if S+C+B \TEXTsymbol{>} 0, Q = Q}%
_{\epsilon _{u}(0)}\text{ = }\frac{2}{3}\text{;}  \label{2/3} \\
\text{for I}_{z} &\text{=}&\text{ 0, if S+C+B \TEXTsymbol{<} 0, Q = Q}%
_{\epsilon _{d}(0)}\text{ = -}\frac{1}{3}\text{.}  \label{- 1/3}
\end{eqnarray}%
There is no quark with I$_{z}$ = 0 and S + C + B = 0.\ \ 

6). Since the most experimental full widths of baryons and mesons are bout
100 Mev, for simplicity, we assume that a fluctuation $\Delta $E of a quark
is

\begin{equation}
\Delta \text{E = 100 S[(1+S}_{Ax}\text{)(J}_{S,}\text{+S}_{Ax}\text{)]}%
\Delta \text{S \ \ \ J}_{S}\text{=}\left\vert \text{S}_{Ax}\right\vert +%
\text{1,2,3, ....}  \label{Dalta-E}
\end{equation}%
The rest mass (m$^{\ast }$) of a quark is the minimum energy of the band.
From (\ref{E(nk)}) and (\ref{Dalta-E}) the rest mass is 
\begin{equation}
\begin{tabular}{l}
$\text{m}^{\ast }\text{ = \{313+ 360 Minimum[(n}_{1}\text{-}\xi \text{)}^{2}%
\text{+(n}_{2}\text{-}\eta \text{)}^{2}\text{+(n}_{3}\text{-}\zeta \text{)}%
^{2}\text{]+}\Delta \text{E+}\Delta \text{\} (Mev)}$ \\ 
\ \ \ \ = m + $\Delta $ \ (Mev),%
\end{tabular}
\label{Rest Mass}
\end{equation}%
This formula (\ref{Rest Mass}) is the united quark mass formula. \ \ \ \ \ \
\ \ \ \ \ \ \ \ 

\section{Deducing Quarks from Energy Bands \ \ \ \ }

From deduced energy bands in Table 1, we can use the above phenomenological
formulae (\ref{IsoSpin})-(\ref{Rest Mass})\ to deduce quarks. For the $%
\Delta $-axis, R = 4, S$_{axis}$ = 0 from (\ref{S-Number}). For single
energy bands, I = 0 from (\ref{IsoSpin}); and S = S$_{axis}$+ $\Delta $S = $%
\Delta $S = $\delta $($\widetilde{n}$) + [1-2$\delta $(S$_{axis}$)]Sign($%
\widetilde{n}$) from (\ref{S+DS}). For $\overrightarrow{n}$ = (0, 0, -2) and
(0, 0, -4), $\Delta $S\ = +1 from (\ref{n/n}) and (\ref{S+DS}); for n = (0,
0, 2), (0, 0, 4) and (0, 0, 6) $\Delta $S = -1 from (\ref{n/n}) and (\ref%
{S+DS}). Using (\ref{Charmed}), (\ref{n/n}) and (\ref{S+DS}), we can find
the charmed number C = +1 when n = (0, 0, -2) and (0, 0, -4). From (\ref{2/3}%
), we can find Q = $\frac{2}{3}$ when n = (0, 0, -2) and (0, 0, -4); from (%
\ref{- 1/3}), Q = - $\frac{1}{3}$ when n = (0, 0, 2), (0, 0, 4) and (0, 0,
6). From (\ref{Dalta-E}) and (\ref{Rest Mass}), we can find the rest masses
(minimE + $\Delta $E). Same flavored quarks with red, yellow or blue colors
have the same rest masses and intrinsic quantum numbers, so we can omit
their colors. We list all results in Table 3:

\ \ \ \ \ \ \ \ \ \ \ \ \ \ \ \ \ \ \ \ \ \ \ \ \ \ \ \ \ \ \ \ \ \ \ \ \ \
\ \ \ \ \ \ \ \ \ \ \ \ \ \ 

\begin{tabular}{|l|}
\hline
\ \ \ \ \ \ \ \ \ \ Table 3. The u$_{C}$(m$^{\ast }$)-quarks and the d$_{S}$%
(m$^{\ast }$)-quarks on the $\Delta $-axis \\ \hline
S$_{axis}$ = 0, I = 0, S = $\Delta $S = $\delta $($\widetilde{n}$) + [1-2$%
\delta $(S$_{axis}$)]Sign($\widetilde{n}$), $\widetilde{n}\text{ }\equiv 
\frac{\text{n}_{1}\text{+n}_{2}\text{+n}_{3}}{\left\vert \text{n}%
_{1}\right\vert \text{+}\left\vert \text{n}_{2}\right\vert \text{+}%
\left\vert \text{n}_{3}\right\vert }$ \\ \hline
\begin{tabular}{|l|l|l|l|l|l|l|l|l|l|l|}
\hline
$\text{n}_{1,}\text{n}_{2,}\text{n}_{3}$ & $\text{E}_{Point}$ & Min. E & $%
\Delta \text{S}$ & \ \ J & I & S & C & Q & $\Delta \text{E}$ & $q_{\text{Name%
}}(m^{\ast })$ \\ \hline
$\text{{\small 0,\ \ 0, \ 0}}$ & $\text{E}_{\Gamma }\text{=0}$ & 313 & 0 & J$%
_{\Gamma }\text{=0}$ & $\frac{1}{2}$ & 0 & 0 & $\frac{2}{3}$ & 0 & $\text{%
u(313+}\Delta \text{)}$ \\ \hline
$\text{{\small 0, \ 0, \ 2}}$ & $\text{E}_{H}\text{=1}$ & 673 & -1 & J$_{%
\text{H}}\text{=1}$ & 0 & -1 & 0 & $\frac{-1}{3}$ & 100 & $\text{d}_{S}\text{%
(773+}\Delta \text{)}$ \\ \hline
$\text{{\small 0, \ 0, -2}}$ & $\text{E}_{\Gamma }\text{=4}$ & 1753 & +1 & J$%
_{\Gamma }\text{=1}$ & 0 & 0 & 1 & $\frac{2}{3}$ & 0 & $\text{u}_{C}\text{%
(1753+}\Delta \text{)}$ \\ \hline
$\text{{\small 0, \ 0, \ 4}}$ & $\text{E}_{H}\text{=9}$ & 3553 & -1 & J$_{%
\text{H}}\text{=2}$ & 0 & -1 & 0 & $\frac{-1}{3}$ & 200 & $\text{d}_{S}\text{%
(3753+}\Delta \text{)}$ \\ \hline
$\text{{\small 0, \ 0, -4}}$ & $\text{E}_{\Gamma }\text{=16}$ & 6073 & +1 & J%
$_{\Gamma }\text{=2}$ & 0 & 0 & 1 & $\frac{2}{3}$ & 0 & $\text{u}_{C}\text{%
(6073+}\Delta \text{)}$ \\ \hline
$\text{{\small 0, \ 0, \ 6}}$ & $\text{E}_{H}\text{=25}$ & $\text{9313}$ & -1
& J$_{\text{H}}\text{=3}$ & 0 & -1 & 0 & $\frac{-1}{3}$ & 300 & $\text{d}_{S}%
\text{(9613+}\Delta \text{)}$ \\ \hline
\end{tabular}
\\ \hline
\end{tabular}

\ \ \ \ \ \ \ \ \ \ \ \ \ \ \ \ \ \ \ \ \ \ \ \ \ \ 

Similarly, for the $\Sigma $-axis, S$_{axis}$ = -2 from (\ref{S-Number}).
For single energy bands, I = 0 from (\ref{IsoSpin}). From (\ref{S+DS}), S = S%
$_{axis}$+ $\Delta $S = -2+ $\Delta $S; the $\Delta $S = $\delta $($%
\widetilde{n}$) + [1-2$\delta $(S$_{axis}$)]Sign($\widetilde{n}$). For $%
\overrightarrow{n}$ = (1, 1, 0), (2, 2, 0), (3, 3, 0) and (4, 4, 0), $\Delta 
$S\ = +1\ from (\ref{n/n}) and (\ref{S+DS}); for $\overrightarrow{n}$ =\
(-1, -1, 0), (-2, -2, 0) and (-3, -3, 0), $\Delta $S\ = -1 from (\ref{n/n})
and (\ref{S+DS}). Using (\ref{Battom}), (\ref{n/n}), (\ref{S+DS}) and (\ref%
{Dalta-E}), we can find the bottom number B = -1 when $\overrightarrow{n}$ =
(3, 3, 0) and (4, 4, 0). From (\ref{2/3}) and (\ref{- 1/3}),we can find the
electric charge Q = -$\frac{1}{3}$ for all quarks. From (\ref{Dalta-E}) and (%
\ref{Rest Mass}), we can deduce rest masses (Min. E + $\Delta $E) of quarks
from the energy bands in Table 2. Same flavored quarks with red, yellow or
blue colors have the same rest masses and intrinsic quantum numbers, so we
omit their colors. We list all results in Table 4:

\ \ \ \ \ \ \ \ \ \ \ \ \ \ \ \ \ \ \ \ \ \ \ \ 

\begin{tabular}{|l|}
\hline
\ \ \ \ \ \ \ \ \ Table 4. The d$_{b}$(m$^{\ast }$), d$_{S}$(m$^{\ast }$)
and d$_{\Omega }$(m$^{\ast }$) Quarks of the $\Sigma $-Axis \\ \hline
{\small S}$_{axis}${\small \ = -2, I = 0, S = S}$_{axis}+\Delta ${\small S = 
}$\delta ${\small (}$\widetilde{n}${\small ) + [1-2}$\delta ${\small (S}$%
_{axis}${\small )]Sign(}$\widetilde{n}${\small ), }$\widetilde{n}\text{ }%
\equiv \frac{\text{n}_{1}\text{+n}_{2}\text{+n}_{3}}{\left\vert \text{n}%
_{1}\right\vert \text{+}\left\vert \text{n}_{2}\right\vert \text{+}%
\left\vert \text{n}_{3}\right\vert }$ \\ \hline
$%
\begin{tabular}{|l|l|l|l|l|l|l|l|l|l|l|}
\hline
{\small E}$_{Point}$ & $\text{\ n}_{1}{\small ,}\text{n}_{2}{\small ,}\text{n%
}_{3}$ & ${\small \Delta }\text{S}$ & {\small S} & {\small B} & {\small Q} & 
${\small \ \ }\text{J}$ & {\small I} & {\small E}$_{\text{Min.}}$ & ${\small %
\Delta }\text{E}$ & $\text{q}_{Name}\text{(m}^{\ast }\text{)}$ \\ \hline
$\text{E}_{\Gamma }\text{=0}$ & {\small (0, 0, 0)} & {\small +2}$^{\#}$ & 
{\small \ 0} & {\small 0} & $\frac{-1}{3}$ & {\small J}$_{\Gamma }\text{ =0}$
& $\frac{1}{2}$ & {\small 313} & {\small 0} & {\small d(313}$\text{+}\Delta $%
{\small )} \\ \hline
$\text{E}_{N}\text{=}\frac{\text{1}}{2}$ & ${\small (}\text{1, 1, 0}{\small )%
}$ & {\small +1} & {\small -1} & {\small 0} & $\frac{-1}{3}$ & {\small J}$_{%
\text{N}}\text{ =1}$ & {\small 0} & $\text{493}$ & {\small 0} & $\text{d}_{S}%
\text{(493+}{\small \Delta }\text{)}$ \\ \hline
$\text{E}_{\Gamma }\text{=2}$ & ${\small (}\text{-1,-1,0}{\small )}$ & 
{\small -1} & {\small -3} & {\small 0} & $\frac{-1}{3}$ & {\small J}$%
_{\Gamma }\text{ =1}$ & {\small 0} & $\text{1033}$ & {\small 0} & $\text{d}%
_{\Omega }\text{(1033+}{\small \Delta }\text{)}$ \\ \hline
$\text{E}_{N}\text{=}\frac{\text{9}}{2}$ & ${\small (}\text{2, 2, 0}{\small )%
}$ & {\small +1} & {\small -1} & {\small 0} & $\frac{-1}{3}$ & {\small J}$_{%
\text{N}}\text{ =2}$ & {\small 0} & $\text{1933}$ & {\small 0} & $\text{d}%
_{S}\text{(1933+}{\small \Delta }\text{)}$ \\ \hline
$\text{E}_{\Gamma }\text{=8}$ & ${\small (}\text{-2,-2,0}{\small )}$ & 
{\small -1} & {\small -3} & {\small 0} & $\frac{-1}{3}$ & {\small J}$%
_{\Gamma }\text{ =2}$ & {\small 0} & $\text{3193}$ & {\small 0} & $\text{d}%
_{\Omega }\text{(3193+}{\small \Delta }\text{)}$ \\ \hline
$\text{E}_{N}\text{=}\frac{\text{25}}{2}$ & ${\small (}\text{3, 3, 0}{\small %
)}$ & {\small +1} & {\small \ 0} & {\small -1} & $\frac{-1}{3}$ & {\small J}$%
_{\text{N}}\text{ =3}$ & {\small 0} & $\text{4813}$ & {\small 100} & $\text{d%
}_{b}\text{(4913+}{\small \Delta }\text{)}$ \\ \hline
$\text{E}_{\Gamma }\text{=18}$ & ${\small (}\text{-3,-3,0}{\small )}$ & 
{\small -1} & {\small -3} & {\small 0} & $\frac{-1}{3}$ & {\small J}$%
_{\Gamma }\text{ =3}$ & {\small 0} & $\text{6793}$ & {\small -300} & $\text{d%
}_{\Omega }\text{(6493+}{\small \Delta }\text{)}$ \\ \hline
$\text{E}_{N}\text{=}\frac{\text{49}}{2}$ & ${\small (}\text{4, 4, 0}{\small %
)}$ & {\small +1} & {\small \ 0} & {\small -1} & $\frac{-1}{3}$ & {\small J}$%
_{\text{N}}\text{ =4}$ & {\small 0} & $\text{9133}$ & {\small 200} & $\text{d%
}_{b}\text{(9333+}{\small \Delta }\text{)}$ \\ \hline
\end{tabular}%
$ \\ \hline
$\ \ \ \ ^{\#}$For ($\text{n}_{1}\text{, n}_{2}\text{, n}_{3}$) = (0, 0, 0), 
$\Delta $S = - S$_{axis}$= +2 from (\ref{DaltaS=-Sax}) \\ \hline
\end{tabular}

\ \ \ \ \ \ \ \ \ \ \ \ \ \ \ \ \ \ \ \ \ \ 

From Tables 3 and 4, we can find that: The unflavored ground quarks are
u(313+$\Delta $) and d(313+$\Delta $). The strange quarks d$_{s}$(493), d$%
_{s}$(773), d$_{s}$(1933), d$_{S}$(3753) d$_{s}$(9613), $\text{d}_{\Omega }%
\text{(1033+}\Delta \text{), d}_{\Omega }\text{(3193+}\Delta \text{) and d}%
_{\Omega }\text{(6493+}\Delta \text{); the strange ground quark is }$d$_{s}$%
(493). The charmed quarks u$_{c}$(1753) and u$_{c}$(6073); the charmed
ground quark is u$_{c}$(1753). The bottom quarks d$_{b}$(4913) and d$_{b}$%
(9333); the bottom ground quark is d$_{b}$(4913). (in Table 11 of \cite%
{0502091} we have shown all low energy quarks, the five deduced ground
quarks are still the ground quarks of all quarks). The five ground quarks
correspond to the five quarks \cite{Quarks} of the current Quark Model. The
deduced intrinsic quantum numbers (I, S, C, B and Q ) of the five ground
quarks are exactly the same as the five current quarks as shown in Table 5A:

\ \ \ \ \ \ \ \ \ \ \ \ \ \ 

\begin{tabular}{l}
\ \ \ \ \ \ \ Table 5A. The Five Deduced Ground Quarks and Current Quarks \\ 
$%
\begin{tabular}{|l|l|l|l|l|l|}
\hline
Quark(m$^{\$}$) & u(313), u & d(313), d & d$_{s}$(493), s & u$_{c}$(1753), c
& d$_{b}$(4913), B \\ \hline
Strange S & 0 \ \ \ \ \ \ \ \ \ 0 & 0 \ \ \ \ \ \ \ \ 0 & -1 \ \ \ \ \ \ \ \
-1 & 0 \ \ \ \ \ \ \ \ \ \ \ \ 0 & 0 \ \ \ \ \ \ \ \ \ \ \ \ \ 0 \\ \hline
Charmed C & 0 \ \ \ \ \ \ \ \ \ 0 & 0 \ \ \ \ \ \ \ \ 0 & 0 \ \ \ \ \ \ \ \
\ 0 & 1 \ \ \ \ \ \ \ \ \ \ \ \ 1 & 0 \ \ \ \ \ \ \ \ \ \ \ \ \ 0 \\ \hline
Bottom B & 0 \ \ \ \ \ \ \ \ \ 0 & 0 \ \ \ \ \ \ \ \ 0 & 0 \ \ \ \ \ \ \ \ \
0 & 0 \ \ \ \ \ \ \ \ \ \ \ \ 0 & -1 \ \ \ \ \ \ \ \ \ \ \ -1 \\ \hline
Isospin I & $\frac{1}{2}$ \ \ \ \ \ \ \ \ $\frac{1}{2}$ & $\frac{1}{2}$ \ \
\ \ \ \ \ \ $\frac{1}{2}$ & 0 \ \ \ \ \ \ \ \ \ 0 & 0 \ \ \ \ \ \ \ \ \ \ \
\ 0 & 0 \ \ \ \ \ \ \ \ \ \ \ \ \ 0\  \\ \hline
I$_{Z}$ & $\frac{1}{2}$ \ \ \ \ \ \ \ \ $\frac{1}{2}$ & -$\frac{1}{2}$ \ \ \
\ \ -$\frac{1}{2}$ & 0 \ \ \ \ \ \ \ \ \ 0 & $0$ \ \ \ \ \ \ \ \ \ \ \ \ 0\
\  & 0 \ \ \ \ \ \ \ \ \ \ \ \ \ 0 \\ \hline
Electric $\text{Q}_{q}$ & $\frac{2}{3}$ \ \ \ \ \ \ \ \ $\frac{2}{3}$ & -$%
\frac{1}{3}$ \ \ \ \ \ -$\frac{1}{3}$ & -$\frac{1}{3}$ \ \ \ \ \ \ -$\frac{1%
}{3}$\  & $\frac{2}{3}$ \ \ \ \ \ \ \ \ \ \ \ $\frac{2}{3}$ & -$\frac{1}{3}$
\ \ \ \ \ \ \ \ \ \ -$\frac{1}{3}$ \\ \hline
\end{tabular}
\ \ $ \\ 
\ \ \ $^{\$}$\ The rest mass of a quark m$^{\ast }$= m + $\Delta \
\rightarrow $ m = m$^{\ast }$ - $\Delta $%
\end{tabular}

\ \ \ \ \ \ \ \ \ \ \ \ \ \ \ \ \ \ \ \ 

The deduced rest masses of the five ground quarks are roughly a constant
(about 390 Mev) larger than the masses of the current quarks, as shown in
Table 5B.

\ \ \ \ \ \ \ \ \ \ \ \ \ \ \ \ \ \ \ \ \ \ \ \ \ \ \ \ \ \ \ \ \ \ \ `\ \ \
\ \ \ \ \ \ \ \ \ \ \ \ \ \ \ \ \ \ 

\begin{tabular}{l}
\ Table 5B Comparing the Rest Masses of Deduced and Current Quarks \\ 
$%
\begin{tabular}{|l|l|l|l|l|l|}
\hline
Quark & Up & Down & Strange & Charmed & bottom \\ \hline
Current Quark(m) & u(2.8) & d(6) & s(105) & c(1225) & b(4500) \\ \hline
Current quark mass & {\small 1.5 to 4} & {\small 4 to 8} & {\small 80 to 130}
& {\small 1250 to 1350} & 
\begin{tabular}{l}
{\small 4.1 to 4.4 G.} \\ 
{\small 4.6 to 4.9 G.}%
\end{tabular}
\\ \hline
Deduced Quark (mass) & u(313) & d(313) & d$_{S}$(493) & u(1753) & d$_{b}$%
(4913) \\ \hline
$\left\vert \text{m}_{Cur.}\text{-m}_{Ded.}\right\vert $ & 310 & 307 & 388 & 
528 & 413 \\ \hline
\end{tabular}%
\ \ $%
\end{tabular}

\ \ \ \ \ \ \ \ \ \ \ \ \ \ \ \ \ \ \ \ \ \ \ \ \ 

These mass differences may originate from different energy reference
systems. If we use the same energy reference system, the deduced masses of
ground quarks will be roughly consistent with the masses of the
corresponding current quarks. Of course, the ultimate test is whether or not
the rest masses of the baryons and mesons composed of the deduced quarks are
consistent with experimental results.

We will deduce the baryons and mesons composed of the quarks in Tables 3 and
4.

\section{The Baryons of the Quarks in Tables 3 and 4}

According to the Quark Model \cite{Quark Model}, a colorless baryon is
composed of three quarks with different\ colors. From Tables 3 and 4, we can
see that there is a term $\Delta \ $of the rest masses in each quark. $%
\Delta $ is a very large unknown constant. Since the rest masses of the
quarks in a baryon are very large (from $\Delta $) and the rest mass of the
baryon composed by three quarks is not, we infer that there will be a strong
binding energy (E$_{Bind}$ = - 3$\Delta $) to cancel 3$\Delta $ from the
three quarks: 
\begin{equation}
\begin{tabular}{l}
$M_{\text{B}}\text{ }\text{= m}_{q_{1}}^{\ast }\text{{\small +\ m}}%
_{q_{2}}^{\ast }\text{{\small \ +m}}_{q_{3}}^{\ast }\text{{\small -}}%
\left\vert \text{E}_{Bind}\right\vert \text{.}$ \\ 
$\text{= }\text{(m}_{q_{1}}\text{+}\Delta \text{)+(m}_{q_{_{N}(313)}}\text{+}%
\Delta \text{)+(m}_{q_{_{N}(313)}}\text{+}\Delta \text{)-3}\Delta .$ \\ 
$\text{= }\text{m}_{q_{1}}\text{+ m}_{q_{_{N}(313)}}\text{+m}_{q_{_{N}(313)}}%
\text{.}$%
\end{tabular}
\label{B-MASS}
\end{equation}%
\ Thus we will omit the term $\Delta \ $in the quark masses and the term $%
-3\Delta \ $in the binding energy from now on. For simplicity's sake, we
only deduced baryons composed of at least two free excited quark q$_{N}$%
(313) (u(313), d(313)) since other baryons have much lower possibilities.
For these baryons, sum laws are:

\begin{equation}
\begin{tabular}{l}
baryon strange number $\ \text{S}_{\text{B}}\text{ }\text{= S}_{q_{1}}\text{%
+ S}_{q_{_{N}(313)}}\text{+ S}_{q_{_{N}(313)}}=\text{ S}_{q_{1(m)}}\text{,}$
\\ 
baryon charmed number $\text{C}_{\text{B\ }}\text{ }\text{= C}_{q_{1}}\text{
+ C}_{q_{_{N}(313)}}\text{ + C}_{q_{_{N}(313)}}\text{= C}_{q_{1}(m)}\text{,}$
\\ 
baryon bottom number $\ $B$_{\text{B}}\text{=}\text{ B}_{q_{1}}\text{+ B}%
_{q_{_{N}(313)}}\text{+ B}_{q_{_{N}(313)}}\text{= B}_{q_{1(m)}}\text{,}$ \\ 
baryon electric charge $\ \text{Q}_{\text{B}}\text{ = Q}_{q_{1}}\text{+ Q}%
_{q_{_{N}(313)}}\text{+ Q}_{_{q_{_{N}(313)}}},$ \\ 
baryon mass M$_{\text{B}}$ = \ $\text{m}_{q_{1}}\text{+ m}_{q_{_{N}(313)}}%
\text{+ m}_{_{q_{_{N}(313)}}}($except charmed baryons) \\ 
charmed baryon M$_{\text{B}}$ = \ $\text{m}_{q_{1}}\text{+ m}_{q_{_{N}(313)}}%
\text{+ m}_{_{q_{_{N}(313)}}}$+ $\Delta e$,%
\end{tabular}
\label{SumB}
\end{equation}
\begin{equation}
\text{where\ \ }\Delta e\text{ = 100C(2I-1), \ \ }  \label{Ebin of B}
\end{equation}
where C is the charmed number and I is the isospin of the baryons.

Using sum laws (\ref{SumB}) and (\ref{Ebin of B}), we can deduce the rest
masses and the intrinsic quantum numbers of baryons from the quarks in
Tables 3 and 4, as shown in Table 6.

\ \ \ \ \ \ \ \ \ \ \ \ \ \ \ \ \ \ \ \ \ \ \ \ \ 

\begin{tabular}{l}
\ \ \ \ \ \ \ \ \ \ Table 6. The Baryons of the Quarks in Table 3 and Table 4
\\ 
\begin{tabular}{|l|l|l|l|l|l|l|l|l|l|l|l|}
\hline
{\small q}$_{i}^{I_{z}}$ & {\small q}$_{j}$ & {\small q}$_{k}$ & {\small I}
& {\small S} & {\small C} & {\small B} & {\small Q} & {\small M} & {\small %
Baryon} & {\small Exper.} & $\frac{\Delta M}{M}${\small \%} \\ \hline
{\small u}$^{\frac{1}{2}}${\small (313)} & {\small u} & {\small d} & $\frac{1%
}{2}$ & {\small 0} & {\small 0} & {\small 0} & {\small 1} & {\small 939} & 
{\small p(939)} & {\small p(938)} & {\small 0.11} \\ \hline
{\small d}$^{\frac{-1}{2}}${\small (313)} & {\small u} & {\small d} & $\frac{%
1}{2}$ & {\small 0} & {\small 0} & {\small 0} & {\small 0} & {\small 939} & 
{\small n(939)} & {\small n(940)} & {\small 0.11} \\ \hline
{\small d}$_{s}^{0}${\small (493)} & {\small u} & {\small d} & {\small 0} & 
{\small -1} & {\small 0} & {\small 0} & {\small 0} & {\small 1119} & $%
\Lambda ${\small (1119)} & $\Lambda ^{0}${\small (1116)} & {\small 0.27} \\ 
\hline
{\small u}$_{c}^{0}${\small (1753)} & {\small u} & {\small d} & {\small 0} & 
{\small 0} & {\small 1} & {\small 0} & {\small 1} & {\small 2279} & $\Lambda
_{c}${\small (2279)} & $\Lambda _{c}^{+}${\small (2285)} & {\small 0.3.} \\ 
\hline
{\small u}$_{c}^{0}${\small (1753)} & {\small u} & {\small u} & {\small 1} & 
{\small 0} & {\small 1} & {\small 0} & {\small 2} & {\small 2479} & $\Sigma
_{c}^{++}${\small (2479)} & $\Sigma _{c}^{++}${\small (2455)} & {\small 1.0}
\\ \hline
{\small u}$_{c}^{0}${\small (1753)} & {\small u} & {\small d} & {\small 1} & 
{\small 0} & {\small 1} & {\small 0} & {\small 1} & {\small 2479} & $\Sigma
_{c}^{+}${\small (2479)} & $\Sigma _{c}^{+}${\small (2455)} & {\small 1.0}
\\ \hline
{\small u}$_{c}^{0}${\small (1753)} & {\small d} & {\small d} & {\small 1} & 
{\small 0} & {\small 1} & {\small 0} & {\small 0} & {\small 2479} & $\Sigma
_{c}^{-}${\small (2479)} & $\Sigma _{c}^{-}${\small (2455)} & {\small 1.0}
\\ \hline
{\small d}$_{b}^{0}${\small (4913)} & {\small u} & {\small d} & {\small 0} & 
{\small 0} & {\small 0} & {\small -1} & {\small 0} & {\small 5539} & $%
\Lambda _{b}${\small (5539)} & $\Lambda _{b}^{0}${\small (5624)} & {\small %
1.5} \\ \hline
{\small d}$_{S}^{0}${\small (773)} & {\small u} & {\small d} & {\small 0} & 
{\small -1} & {\small 0} & {\small 0} & {\small 0} & {\small 1399} & $%
\Lambda ${\small (1399)} & $\Lambda ${\small (1405)} & {\small 0.4} \\ \hline
{\small d}$_{S}^{0}${\small (1933)} & {\small u} & {\small d} & {\small 0} & 
{\small -1} & {\small 0} & {\small 0} & {\small 0} & {\small 2559} & $%
\Lambda ${\small (2559)} & $\Lambda ${\small (2585)}$^{\ast \ast }$ & 
{\small 1.0} \\ \hline
{\small d}$_{S}^{0}${\small (3753)} & {\small u} & {\small d} & {\small 0} & 
{\small -1} & {\small 0} & {\small 0} & {\small 0} & {\small 4375} & $%
\Lambda ${\small (4375)} & {\small Prediction} &  \\ \hline
{\small d}$_{S}^{0}${\small (9613)} & {\small u} & {\small d} & {\small 0} & 
{\small -1} & {\small 0} & {\small 0} & {\small 0} & {\small 10239} & $%
\Lambda ${\small (10239)} & {\small Prediction} &  \\ \hline
{\small u}$_{c}^{0}${\small (6073)} & {\small u} & {\small d} & {\small 0} & 
{\small 0} & {\small 1} & {\small 0} & {\small 1} & {\small 6599} & $\Lambda
_{C}^{+}${\small (6599)} & {\small Prediction} &  \\ \hline
{\small d}$_{b}^{0}${\small (9333)} & {\small u} & {\small d} & {\small 0} & 
{\small 0} & {\small 0} & {\small -1} & {\small 0} & {\small 9959} & $%
\Lambda _{b}^{0}${\small (9959)} & {\small Prediction} &  \\ \hline
{\small d}$_{\Omega }^{0}${\small (1033)} & {\small d} & {\small d} & 
{\small 0} & {\small -3} & {\small 0} & {\small 0} & {\small -1} & {\small %
1659} & $\Omega ^{-}${\small (1659)} & $\Omega ^{-}${\small (1672)} & 
{\small 0.8} \\ \hline
\end{tabular}
\\ 
\ \ In the Table, u $\equiv $ {\small u}$^{\frac{1}{2}}${\small (313) and d} 
$\equiv $ {\small d}$^{\frac{-1}{2}}${\small (313).}%
\end{tabular}

\ \ \ \ \ \ \ \ \ \ \ \ \ \ \ 

Table 6 shows that the deduced intrinsic quantum numbers ( I, S, C, B and Q)
of the baryons match experimental results \cite{Baryon04} exactly and that
the deduced rest masses of the baryons are consistent with more than 98.5\%
of experimental results.

\section{The Mesons of the Quarks in Tables 3 and 4}

According to the Quark Model \cite{Quark Model}, a colorless meson is
composed of a quark q$_{i}$ with a color and an antiquark $\overline{q_{j}}$
with the anticolor of q$_{i}$. For the same flavor, the three pairs of
colored quark and antiquark (q$_{i}^{\func{Re}d}\overline{q_{j}^{\func{Re}d}%
\text{ \ }}$= q$_{i}^{Yellow}\overline{q_{j}^{Yellow}\text{ }}$=q$_{i}^{Blue}%
\overline{q_{j}^{Blue}\text{ }}$) have the same rest masses and intrinsic
quantum numbers (I, S, C, B, Q).\ Thus we can omit the color when we deduce
the rest masses and intrinsic quantum numbers of the mesons. For mesons, the
sum laws are

\begin{equation}
\begin{tabular}{l}
meson strange number $\text{S}_{\text{M}}\text{ }\text{= S}_{q_{i}}\text{+ S}%
_{\overline{q_{j}}}\text{,}$ \\ 
meson charmed number $\text{C}_{\text{M\ }}\text{ }\text{= C}_{q_{i}}\text{
+ C}_{\overline{q_{j}}}\text{ ,}$ \\ 
meson bottom number B$_{\text{M}}\text{=}\text{ B}_{q_{i}}\text{+ B}_{%
\overline{q_{j}}}\text{,}$ \\ 
meson electric charge $\text{Q}_{\text{M}}\text{ = Q}_{q_{i}}\text{+ Q}_{%
\overline{q_{j}}}.$%
\end{tabular}
\label{Sum(SCbQ)}
\end{equation}
There is a strong interaction between the quark and antiquark (colors), but
we do not know how large it is. Since the rest masses of the quark and
antiquark in mesons are large (from $\Delta $) and the rest mass of the
meson composed of the quark and antiquark is not, we infer that there will
be a large portion of binding energy (- 2$\Delta $) to cancel 2$\Delta $
from the quark and antiquark and a small amount of binding energy as shown
in the following 
\begin{equation}
\text{E}_{B}\text{(q}_{i}\overline{q_{j}}\text{) = -2}\Delta \text{ - 337 +
100[}\frac{\Delta \text{m}}{\text{m}_{g}}\text{ +DS -}\ \widetilde{m}\text{+}%
\gamma \text{(i,j) -2I}_{i}\text{I}_{j}\text{]}  \label{Eb-Meson}
\end{equation}
where $\Delta $ = $\frac{1}{3}\left\vert \text{E}_{bind}\right\vert $ (\ref%
{313&Dalta}) is $\frac{1}{3}$ of the baryon binding energy (an unknown large
constant, $\Delta $ \TEXTsymbol{>}\TEXTsymbol{>} m$_{\text{P}}$= 938 Mev), $%
\Delta $m = $\left\vert \text{m}_{i}\text{-m}_{j}\right\vert $, DS =$%
\left\vert \text{(}\Delta \text{S)}_{i}\text{- (}\Delta \text{S)}%
_{j}\right\vert $. m$_{g}$ = 939 (Mev) unless 
\begin{equation}
\begin{tabular}{|l|l|l|l|}
\hline
m$_{i}$(or m$_{j}$) equals & m$_{C}\geqslant $ 6073 & m$_{b}\geqslant $ 9333
& m$_{S}\geqslant $ 9613 \\ \hline
$\ \ \ \text{m}_{g}${\small \ will equal to} & 1753(Table 4) & 4913 (Table7)
& 3753(Table 4). \\ \hline
\end{tabular}
\label{m(g)}
\end{equation}

$\ \widetilde{m}$ = $\frac{m_{i}\times m_{j}}{\text{m}_{g_{i}}\times \text{m}%
_{g_{j}}}$ \ \ m$_{g_{i}}$ = m$_{g_{j}}^{\text{ \ }}$ = 939 (Mev) unless 
\begin{equation}
\begin{tabular}{|l|l|l|l|l|l|}
\hline
\ m$_{i}$(or m$_{j}$) = & m$_{q_{_{N}}}$=313 & m$_{d_{s}}$=493 & m$%
_{u_{c}}\succeq $1753 & m$_{d_{S}}$\TEXTsymbol{>} 3753, & m$_{d_{b}}\succeq $
4913 \\ \hline
\ m$_{g_{j}}$ (or m$_{g_{j}}^{\text{ \ }})$ & 313 & 493 & 1753 & 3753, & 
4913. \\ \hline
\end{tabular}
\label{M(gi)}
\end{equation}
\ If\ q$_{i\text{ }}$and q$_{j}$ are both ground quarks, $\gamma $(i, j) =
0. If\ q$_{i\text{ }}$and q$_{j}$ are not both ground quarks, for q$_{i\text{
}}$= q$_{j}$, $\gamma $(i, j) = -$1$; for q$_{i\text{ }}\neq $ q$_{j}$,$\
\gamma $(i, j) = +1. S$_{i}$ (or S$_{j}$) is the strange number of the quark
q$_{i}$ (or q$_{j}$). I$_{i}$ (or I$_{j}$) is the isospin of the quark q$%
_{i} $ (or q$_{j}$). When we deduce rest masses of mesons, we will omit the $%
\Delta $ part in the quark and antiquark mass and omit "- 2 $\Delta "$%
binding energy (\ref{Eb-Meson}).

From the quarks in Tables 3 and 4, we can use (\ref{Sum(SCbQ)}) and (\ref%
{Eb-Meson}) to deduce the rest masses and the intrinsic\ quantum numbers (I,
S, C, B and Q) of mesons as shown in Table 7.

\ \ \ \ \ \ \ \ \ \ \ \ \ \ \ \ \ \ \ \ \ \ \ \ \ \ \ \ \ 

\begin{tabular}{l}
$\ \ \ \ \ \ \ \ \ \ \ \text{Table 7.\ \ The Deduced Mesons of the Quarks in
Tables 3 and 4}$ \\ 
$%
\begin{tabular}{|l|l|l|l|l|l|l|l|}
\hline
$\text{q}_{i}^{\Delta S}\text{(m}_{i}\text{)}{\small \ }\overline{\text{q}%
_{j}^{\Delta S}\text{(m}_{j}\text{)}}$ & $\frac{100\Delta \text{m}}{\text{939%
}}$ & {\small DS} & {\small -100}$\widetilde{m}$ & {\small E}$_{bind}$ & 
{\small Deduced} & {\small Experiment} & {\small R} \\ \hline
$\text{q}_{N}^{0}${\small (313)}$\overline{_{N}^{0}\text{(313)}}$ & {\small 0%
} & ${\small 0}$ & {\small -100} & {\small - 487}$^{\#}$ & $\pi ${\small %
(139)} & $\pi ${\small (138)} & {\small 0.7} \\ \hline
{\small q}$_{N}^{0}${\small (313)}$\overline{\text{d}_{S}^{1}\text{(493)}}$
& {\small 19} & {\small 1} & {\small -100} & {\small - 318} & {\small K(488)}
& {\small K(494)} & {\small 0.2} \\ \hline
{\small d}$_{S}^{1}${\small (493)}$\overline{\text{d}_{S}^{1}\text{(493)}}$
& {\small 0} & {\small 0} & {\small -100} & {\small - 437} & $\eta ${\small %
(549)} & $\eta ${\small (548)} & {\small 0.2} \\ \hline
{\small u}$_{C}^{1}${\small (1753)}$\overline{_{N}^{0}\text{(313)}}$ & 
{\small 153} & {\small 1} & {\small -100} & {\small - 184} & {\small D(1882)}
& {\small D(1869)} & {\small 0.7} \\ \hline
{\small u}$_{C}^{1}${\small (1753)}$\overline{\text{q}_{S}^{1}\text{(493)}}$
& {\small 134} & {\small 0} & {\small -100} & {\small - 303} & {\small D}$%
_{S}${\small (1943)} & {\small D}$_{S}${\small (1969)} & {\small 0.4} \\ 
\hline
{\small u}$_{C}^{1}${\small (1753)}$\overline{\text{u}_{C}^{1}\text{(1753)}}$
& {\small 0} & {\small 0} & {\small -100} & {\small -437} & {\small J/}$\psi 
${\small (3069)} & {\small J/}$\psi ${\small (3097)} & {\small 0.9} \\ \hline
$\text{q}_{N}^{0}${\small (313)}$\overline{\text{d}_{b}^{1}\text{(4913)}}$ & 
{\small 490} & {\small 1} & {\small -100} & {\small \ 153} & {\small B(5379)}
& {\small B(5279)} & {\small 1.9} \\ \hline
{\small d}$_{S}^{1}${\small (493)}$\overline{\text{d}_{b}^{1}\text{(4913)}}$
& {\small 471} & ${\small 0}$ & {\small -100} & {\small \ 34} & {\small B}$%
_{S}${\small (5440)} & {\small B}$_{S}${\small (5370)} & {\small 1.3} \\ 
\hline
{\small u}$_{C}^{1}${\small (1753)}$\overline{\text{d}_{b}^{1}\text{(4913)}}$
& {\small 337} & {\small 0} & {\small -100} & {\small -100} & {\small B}$%
_{C} ${\small (6566)} & {\small B}$_{C}${\small (6400)} & {\small 2.6} \\ 
\hline
{\small d}$_{b}^{1}${\small (4913)}$\overline{\text{d}_{b}^{1}\text{(4913)}}$
& {\small 0} & {\small 0} & {\small -100} & {\small -\ 437} & $\Upsilon $%
{\small (9389)} & $\Upsilon ${\small (9460)} & {\small 0.8} \\ \hline
{\small d}$_{S}^{-1}${\small (773)}$\overline{\text{d}_{S}^{-1}\text{(773)}}$
& {\small 0} & {\small 0} & {\small -68} & {\small -505} & $\eta ${\small %
(1041)} & ${\small \phi (1020)}$ & {\small 2.0} \\ \hline
{\small d}$_{S}^{1}${\small (1933)}$\overline{\text{d}_{S}^{1}\text{(1933)}}$
& {\small 0} & {\small 0} & {\small -424} & {\small -861} & $\eta ${\small %
(3005)} & $\eta _{c}${\small (2980)} & {\small 0.8} \\ \hline
{\small d}$_{S}^{1}${\small (9333)}$\overline{\text{d}_{S}^{1}\text{(9333)}}$
& {\small 0} & {\small 0} & {\small -361} & {\small -798} & $\Upsilon $%
{\small (17868)} & {\small prediction} &  \\ \hline
{\small u}$_{C}^{1}${\small (6073)}$\overline{\text{u}_{C}^{1}\text{(6073)}}$
& {\small 0} & {\small 0} & {\small -1200} & {\small -1637} & $\psi ${\small %
(10509)} & $\Upsilon ${\small (10355)} & {\small 1.5} \\ \hline
{\small d}$_{\Omega }^{-1}${\small (1033)}$\overline{\text{d}_{\Omega }^{-1}%
\text{(1033)}}$ & {\small 0} & {\small 0} & {\small -121} & {\small -558} & $%
\eta ${\small (1508)} & {\small f}$_{0}${\small (1507)} & {\small 0.7} \\ 
\hline
{\small q}$_{N}^{0}${\small (313)}$\overline{\text{d}_{S}^{-1}\text{(773)}}$
& {\small 49} & {\small 1} & {\small -82} & {\small -170} & {\small K(916)}
& {\small K(892)} & {\small 2.7} \\ \hline
{\small q}$_{N}^{0}${\small (313)}$\overline{\text{d}_{S}^{1}\text{(1933)}}$
& {\small 171} & {\small 1} & {\small -206} & {\small -170} & {\small K(2076)%
} & {\small K}$_{4}^{\ast }${\small (2045)} & {\small 1.5} \\ \hline
{\small q}$_{N}^{0}${\small (313)}$\overline{\text{d}_{S}^{-1}\text{(3753)}}$
& {\small 347} & {\small 1} & {\small -400} & {\small -190} & {\small K(3876)%
} & {\small prediction} &  \\ \hline
{\small q}$_{N}^{0}${\small (313)}$\overline{\text{d}_{S}^{-1}\text{(9613)}}$
& {\small 248} & {\small 1} & {\small -256} & {\small -145} & {\small K(9781)%
} & {\small prediction} &  \\ \hline
{\small q}$_{N}^{0}${\small (313)}$\overline{\text{d}_{b}^{1}\text{(9333)}}$
& {\small 183} & {\small 1} & {\small -190} & {\small -144} & {\small B(9502)%
} & {\small prediction} &  \\ \hline
{\small u}$_{C}^{1}${\small (6073)}$\overline{\text{q}_{N}^{0}\text{(313)}}$
& {\small 328} & {\small 1} & {\small 346.4} & {\small -155} & {\small %
D(6231)} & {\small prediction} &  \\ \hline
{\small d}$_{S}^{1}${\small (493)}$\overline{\text{d}_{S}^{-1}\text{(773)}}$
& {\small 30} & {\small 2} & {\small 256.1} & {\small -90} & $\eta ${\small %
(1177)} & $\eta ${\small (1170)} & {\small 0.6} \\ \hline
{\small d}$_{S}^{1}${\small (493)}$\overline{\text{d}_{S}^{-1}\text{(3753)}}$
& {\small 347} & {\small 2} & {\small 339.7} & {\small -90} & $\eta ${\small %
(4156)} & $\psi ${\small (4159)} & {\small .07} \\ \hline
{\small d}$_{S}^{1}${\small (493)}$\overline{\text{d}_{S}^{-1}\text{(9613)}}$
& {\small 243} & {\small 2} & {\small 256.1} & {\small -50} & $\eta ${\small %
(10056)} & $\Upsilon ${\small (10023)} & {\small 0.4} \\ \hline
{\small u}$_{C}^{1}${\small (1753)}$\overline{\text{d}_{S}^{-1}\text{(773)}}$
& {\small 104} & {\small 2} & {\small 82.3} & {\small -15} & {\small D}$_{S}$%
{\small (2511)} & {\small D}$_{S_{1}}${\small (2535)} & {\small 1.0} \\ 
\hline
{\small d}$_{S}^{1}${\small (9613)}$\overline{\text{d}_{S}^{-1}\text{(773)}}$
& {\small 235} & {\small 0} & {\small 211} & {\small -212} & $\eta ${\small %
(10174)} & $\chi ${\small (10232)} & {\small 0.6} \\ \hline
\end{tabular}%
\ \ $ \\ 
\ \ \ $^{\#}${\small \ For}$\ ${\small q}$_{N}^{0}${\small (313)}$\overline{%
_{N}^{0}\text{(313)}}\ ,${\small 100}$\times ${\small \ 2I}$_{i}${\small I}$%
_{j}${\small \ = 50 and for other pairs 100}$\times ${\small \ 2I}$_{i}$%
{\small I}$_{j}${\small \ = 0.}%
\end{tabular}

\ \ \ \ \ \ \ \ \ \ \ \ \ \ \ \ \ \ \ \ \ \ \ 

Table 7 shows that the deduced intrinsic quantum numbers match experimental
results \cite{Meson04} exactly. The deduced rest masses are more than 97\%
consistent with experimental results.\ 

\section{Predictions\ \ \ \ \ \ \ \ \ \ \ \ \ \ \ \ \ \ \ \ \ \ \ \ \ \ \ \
\ \ \ \ \ \ \ \ \ \ \ \ \ \ \ \ \ \ \ \ \ \ \ \ \ \ \ \ \ \ \ \ \ \ \ \ \ \
\ \ \ \ \ \ \ \ \ \ \ \ \ \ \ \ \ \ \ \ \ \ \ \ \ \ \ \ \ \ \ \ \ \ \ \ \ \
\ \ \ \ \ \ \ \ \ \ \ \ \ \ \ \ \ \ \ \ \ \ \ \ \ \ \ \ \ \ \ \ \ \ \ \ \ \
\ \ \ \ \ \ \ \ \ \ \ \ \ \ \ \ \ \ \ \ \ \ \ \ \ \ \ \ \ \ \ \ \ \ \ \ \ \
\ \ \ \ \ \ \ \ \ \ \ \ \ \ \ \ \ \ \ \ \ \ \ \ \ \ \ \ \ \ \ \ \ \ \ \ \ \
\ \ \ \ \ \ \ \ \ \ \ \ \ \ \ \ \ \ \ \ \ \ \ \ \ \ \ \ \ \ \ \ \ \ \ \ \ \
\ \ \ \ \ \ \ \ \ \ \ \ \ \ \ \ \ \ \ \ \ \ \ \ \ \ \ \ \ \ \ \ \ \ \ \ \ \
\ \ \ \ \ \ \ \ \ \ \ \ \ \ \ \ \ \ \ \ \ \ \ \ \ \ \ \ \ \ \ \ \ \ \ \ \ \
\ \ \ \ \ \ \ \ \ \ \ \ \ \ \ \ \ \ \ \ \ \ \ \ \ \ \ \ \ \ \ \ \ \ \ \ \ \
\ \ \ \ \ \ \ \ \ \ \ \ \ \ \ \ \ \ \ \ \ \ \ \ \ \ \ \ \ \ \ \ \ \ \ \ \ \
\ \ \ \ \ \ \ \ \ \ \ \ \ \ \ \ \ \ \ \ \ \ \ \ \ \ \ \ \ \ \ \ \ \ \ \ \ \
\ \ \ \ \ \ \ \ \ \ \ \ \ \ \ \ \ \ \ \ \ \ \ \ \ \ \ \ \ \ \ \ \ \ \ \ \ \
\ \ \ \ \ \ \ \ \ \ \ \ \ \ \ \ \ \ \ \ \ \ \ \ \ \ \ \ \ \ \ \ \ \ \ \ \ \
\ \ \ \ \ \ \ \ \ \ \ \ \ \ \ \ \ \ \ \ \ \ \ \ \ \ \ \ \ \ \ \ \ \ \ \ \ \
\ \ \ \ \ \ \ \ \ \ \ \ \ \ \ \ \ \ \ \ \ \ \ \ \ \ \ \ \ \ \ \ \ \ \ \ \ \
\ \ \ \ \ \ \ \ \ \ \ \ \ \ \ \ \ \ \ \ \ \ \ \ \ \ \ \ \ \ \ \ \ \ \ \ \ \
\ \ \ \ \ \ \ \ \ \ \ \ \ \ \ \ \ \ \ \ \ \ \ \ \ \ \ \ \ \ \ \ \ \ \ \ \ \
\ \ \ \ \ \ \ \ \ \ \ \ \ \ \ \ \ \ \ \ \ \ \ \ \ \ \ \ \ \ \ \ \ \ \ \ \ \
\ \ \ \ \ \ \ \ \ \ \ \ \ \ \ \ \ \ \ \ \ \ \ \ \ \ \ \ \ \ \ \ \ \ \ \ \ \
\ \ \ \ \ \ \ \ \ \ \ \ \ \ \ \ \ \ \ \ \ \ \ \ \ \ \ \ \ \ \ \ \ \ \ \ \ \
\ \ \ \ \ \ \ \ \ \ \ \ \ \ \ \ \ \ \ \ \ \ \ \ \ \ \ \ \ \ \ \ \ \ \ \ \ \
\ \ \ \ \ \ \ \ \ \ \ \ \ \ \ \ \ \ \ \ \ \ \ \ \ \ \ \ \ \ \ \ \ \ \ \ \ \
\ \ \ \ \ \ \ \ \ \ \ \ \ \ \ \ \ \ \ \ \ \ \ \ \ \ \ \ \ \ \ \ \ \ \ \ \ \
\ \ \ \ \ \ \ \ \ \ \ \ \ \ \ \ \ \ \ \ \ \ \ \ \ \ \ \ \ \ \ \ \ \ \ \ \ \
\ \ \ \ \ \ \ \ \ \ \ \ \ \ \ \ \ \ \ \ \ \ \ \ \ \ \ \ \ \ \ \ \ \ \ \ \ \
\ \ \ \ \ \ \ \ \ \ \ \ \ \ \ \ \ \ \ \ \ \ \ \ \ \ \ \ \ \ \ \ \ \ \ \ \ \
\ \ \ \ \ \ \ \ \ \ \ \ \ \ \ \ \ \ \ \ \ \ \ \ \ \ \ \ \ \ \ \ \ \ \ \ \ \
\ \ \ \ \ \ \ \ \ \ \ \ \ \ \ \ \ \ \ \ \ \ \ \ \ \ \ \ \ \ \ \ \ \ \ \ \ \
\ \ \ \ \ \ \ \ \ \ \ \ \ \ \ \ \ \ \ \ \ \ \ \ \ \ \ \ \ \ \ \ \ \ \ \ \ \
\ \ \ \ \ \ \ \ \ \ \ \ \ \ \ \ \ \ \ \ \ \ \ \ \ \ \ \ \ \ \ \ \ \ \ \ \ \
\ \ \ \ \ \ \ \ \ \ \ \ \ \ \ \ \ \ \ \ \ \ \ \ \ \ \ \ \ \ \ \ \ \ \ \ \ \
\ \ \ \ \ \ \ \ \ \ \ \ \ \ \ \ \ \ \ \ \ \ \ \ \ \ \ \ \ \ \ \ \ \ \ \ \ \
\ \ \ \ \ \ \ \ \ \ \ \ \ \ \ \ \ \ \ \ \ \ \ \ \ \ \ \ \ \ \ \ \ \ \ \ \ \
\ \ \ \ \ \ \ \ \ \ \ \ \ \ \ \ \ \ \ \ \ \ \ \ \ \ \ \ \ \ \ \ \ \ \ \ \ \
\ \ \ \ \ \ \ \ \ \ \ \ \ \ \ \ \ \ \ \ \ \ \ \ \ \ \ \ \ \ \ \ \ \ \ \ \ \
\ \ \ \ \ \ \ \ \ \ \ \ \ \ \ \ \ \ \ \ \ \ \ \ \ \ \ \ \ \ \ \ \ \ \ \ \ \
\ \ \ \ \ \ \ \ \ \ \ \ \ \ \ \ \ \ \ \ \ \ \ \ \ \ \ \ \ \ \ \ \ \ \ \ \ \
\ \ \ \ \ \ \ \ \ \ \ \ \ \ \ \ \ \ \ \ \ \ \ \ \ \ \ \ \ \ \ \ \ \ \ \ \ \
\ \ \ \ \ \ \ \ \ \ \ \ \ \ \ \ \ \ \ \ \ \ \ \ \ \ \ \ \ \ \ \ \ \ \ \ \ \
\ \ \ \ \ \ \ \ \ \ \ \ \ \ \ \ \ \ \ \ \ \ \ \ \ \ \ \ \ \ \ \ \ \ \ \ \ \
\ \ \ \ \ \ \ \ \ \ \ \ \ \ \ \ \ \ \ \ \ \ \ \ \ \ \ \ \ \ \ \ \ \ \ \ \ \
\ \ \ \ \ \ \ \ \ \ \ \ \ \ \ \ \ \ \ \ \ \ \ \ \ \ \ \ \ \ \ \ \ \ \ \ \ \
\ \ \ \ \ \ \ \ \ \ \ \ \ \ \ \ \ \ \ \ \ \ \ \ \ \ \ \ \ \ \ \ \ \ \ \ \ \
\ \ \ \ \ \ \ \ \ \ \ \ \ \ \ \ \ \ \ \ \ \ \ \ \ \ \ \ \ \ \ \ \ \ \ \ \ \
\ \ \ \ \ \ \ \ \ \ \ \ \ \ \ \ \ \ \ \ \ \ \ \ \ \ \ \ \ \ \ \ \ \ \ \ \ \
\ \ \ \ \ \ \ \ \ \ \ \ \ \ \ \ \ \ \ \ \ \ \ \ \ \ \ \ \ \ \ \ \ \ \ \ \ \
\ \ \ \ \ \ \ \ \ \ \ \ \ \ \ \ \ \ \ \ \ \ \ \ \ \ \ \ \ \ \ \ \ \ \ \ \ \
\ \ \ \ \ \ \ \ \ \ \ \ \ \ \ \ \ \ \ \ \ \ \ \ \ \ \ \ \ \ \ \ \ \ \ \ \ \
\ \ \ \ \ \ \ \ \ \ \ \ \ \ \ \ \ \ \ \ \ \ \ \ \ \ \ \ \ \ \ \ \ \ \ \ \ \
\ \ \ \ \ \ \ \ \ \ \ \ \ \ \ \ \ \ \ \ \ \ \ \ \ \ \ \ \ \ \ \ \ \ \ \ \ \
\ \ \ \ \ \ \ \ \ \ \ \ \ \ \ \ \ \ \ \ \ \ \ \ \ \ \ \ \ \ \ \ \ \ \ \ \ \
\ \ \ \ \ \ \ \ \ \ \ \ \ \ \ \ \ \ \ \ \ \ \ \ \ \ \ \ \ \ \ \ \ \ \ \ \ \
\ \ \ \ \ \ \ \ \ \ \ \ \ \ \ \ \ \ \ \ \ \ \ \ \ \ \ \ \ \ \ \ \ \ \ \ \ \
\ \ \ \ \ \ \ \ \ \ \ \ \ \ \ \ \ \ \ \ \ \ \ \ \ \ \ \ \ \ \ \ \ \ \ \ \ \
\ \ \ \ \ \ \ \ \ \ \ \ \ \ \ \ \ \ \ \ \ \ \ \ \ \ \ \ \ \ \ \ \ \ \ \ \ \
\ \ \ \ \ \ \ \ \ \ \ \ \ \ \ \ \ \ \ \ \ \ \ \ \ \ \ \ \ \ \ \ \ \ \ \ \ \
\ \ \ \ \ \ \ \ \ \ \ \ \ \ \ \ \ \ \ \ \ \ \ \ \ \ \ \ \ \ \ \ \ \ \ \ \ \
\ \ \ \ \ \ \ \ \ \ \ \ \ \ \ \ \ \ \ \ \ \ \ \ \ \ \ \ \ \ \ \ \ \ \ \ \ \
\ \ \ \ \ \ \ \ \ \ \ \ \ \ \ \ \ \ \ \ \ \ \ \ \ \ \ \ \ \ \ \ \ \ \ \ \ \
\ \ \ \ \ \ \ \ \ \ \ \ \ \ \ \ \ \ \ \ \ \ \ \ \ \ \ \ \ \ \ \ \ \ \ \ \ \
\ \ \ \ \ \ \ \ \ \ \ \ \ \ \ \ \ \ \ \ \ \ \ \ \ \ \ \ \ \ \ \ \ \ \ \ \ \
\ \ \ \ \ \ \ \ \ \ \ \ \ \ \ \ \ \ \ \ \ \ \ \ \ \ \ \ \ \ \ \ \ \ \ \ \ \
\ \ \ \ \ \ \ \ \ \ \ \ \ \ \ \ \ \ \ \ \ \ \ \ \ \ \ \ \ \ \ \ \ \ \ \ \ \
\ \ \ \ \ \ \ \ \ \ \ \ \ \ \ \ \ \ \ \ \ \ \ \ \ \ \ \ \ \ \ \ \ \ \ \ \ \
\ \ \ \ \ \ \ \ \ \ \ \ \ \ \ \ \ \ \ \ \ \ \ \ \ \ \ \ \ \ \ \ \ \ \ \ \ \
\ \ \ \ \ \ \ \ \ \ \ \ \ \ \ \ \ \ \ \ \ \ \ \ \ \ \ \ \ \ \ \ \ \ \ \ \ \
\ \ \ \ \ \ \ \ \ \ \ \ \ \ \ \ \ \ \ \ \ \ \ \ \ \ \ \ \ \ \ \ \ \ \ \ \ \
\ \ \ \ \ \ \ \ \ \ \ \ \ \ \ \ \ \ \ \ \ \ \ \ \ \ \ \ \ \ \ \ \ \ \ \ \ \
\ \ \ \ \ \ \ \ \ \ \ \ \ \ \ \ \ \ \ \ \ \ \ \ \ \ \ \ \ \ \ \ \ \ \ \ \ \
\ \ \ \ \ \ \ \ \ \ \ \ \ \ \ \ \ \ \ \ }

This paper predicts some quarks, baryons and mesons shown in the following
list:

\begin{tabular}{|l|l|l|l|}
\hline
{\small q}$_{i}${\small (m)} & {\small Baryon[Exper.]} & {\small q}$_{N}$%
{\small (m)}$\overline{\text{q}_{i}\text{{\small (m)}}}${\small \ [Exper. ]}
& {\small q}$_{i}${\small (m)}$\overline{\text{q}_{i}\text{(m)}}${\small \
[Exper.]} \\ \hline
$\text{u}_{C}${\small (6073)} & $\Lambda _{c}${\small (6599) [ ? \ ]} & 
{\small D(6231) \ \ \ \ \ \ [ ? \ \ \ \ \ \ \ ]} & $\psi ${\small (10509)\ \
\ [}$\Upsilon ${\small (10355)]} \\ \hline
$\text{d}_{S}\text{({\small 773})}$ & $\Lambda ${\small (1399)} [$\Lambda $%
{\small (1406)]} & {\small K(916) }$\ \ \ \ \ \ $[{\small K(892)]} & {\small %
\ }$\eta ${\small (1041)\ \ \ \ \ [}$\phi ${\small (1020) ]} \\ \hline
$\text{d}_{S}\text{({\small 1933})}$ & $\Lambda ${\small (2559) [}$\Lambda $%
{\small (2585)}$^{\ast \ast }$] & {\small K(2076) \ \ \ \ \ [K}$_{4}^{\ast }$%
{\small (2045) ]} & {\small \ }$\eta ${\small (3005)\ \ \ \ \ [}$\eta _{c}$%
{\small (2980)]} \\ \hline
$\text{d}_{b}\text{({\small 9333})}$ & $\Lambda _{b}${\small (9959) [ ? \ ]}
& B(9502) \ \ [ \ \ ?\ \ \ \ ] & $\Upsilon $(17868) \ [\ \ \ ? \ \ \ \ ] \\ 
\hline
\end{tabular}

\ $\Lambda ${\small (2585)}$^{\ast \ast }$ Evidence of existence is only fair

It is very important to pay attention to the $\Upsilon $(3S)-meson (mass m =
10,355.2 $\pm $ 0.4 Mev, full width $\Gamma $\ = 26.3 $\pm $ 3.5 kev). We
compare the mesons J/$\psi $(3097), $\Upsilon $(9460) and $\Upsilon $(10355)
shown as follow list

\begin{tabular}{l}
u$_{C}^{1}$(1753)$\overline{\text{u}_{C}^{1}\text{(1753)}}$ = J/$\psi $%
(3069) \ [J/$\psi $(3096.916$\pm $0.011), $\Gamma $ = 91.0 $\pm $ 3.2kev] \\ 
d$_{b}^{1}$(4913)$\overline{\text{d}_{b}^{1}\text{(4913)}}$ = $\Upsilon $%
(9389) \ \ \ \ \ [$\Upsilon $(9460.30$\pm $0.26), $\ \ \ \ \ \ \ \Gamma $ =
53.0 $\pm $ 1.5kev] \\ 
u$_{C}^{1}$(6073)$\overline{\text{u}_{C}^{1}\text{(6073)}}$ = $\psi $(10509)
\ \ [$\Upsilon $(10.355.2 $\pm $ 0.4), $\ \ \ \ \ \Gamma $\ = 26.3 $\pm $
3.5 kev]%
\end{tabular}

$\Upsilon $(3S) has more than three times larger of a mass than J/$\psi $%
(1S)\ (m = 3096.916 $\pm $ 0.011 Mev)\ and more than three times longer of
lifetime than J/$\psi $(1S)\ (full width $\Gamma $ = 91.0 $\pm $ \ 3.2 kev).
It is well known that the discovery\ of J/$\psi $(1S)\ is also the discovery
of charmed\ quark c (u$_{c}$(1753)) and that the discovery\ of $\Upsilon $%
(9460)\ is also the discovery of bottom\ quark b (d$_{b}$(4913)). Similarly
the discovery of $\Upsilon $(3S) will be the discovery of a very important
new quark---the u$_{C}$(6073)-quark.

\section{Discussion\ \ \ \ \ \ \ \ \ \ \ \ \ \ \ \ \ \ \ \ \ \ \ \ \ \ \ \ \
\ \ \ \ }

1). The fact that physicists have not found any free quark shows that the
binding energies are very large. The baryon binding energy -3$\Delta $
(meson - 2$\Delta $ ) is a phenomenological approximation of the color's
strong interaction energy in a baryon (a meson). The binding energy -3$%
\Delta $ (-2$\Delta $) is always cancelled by the corresponding parts 3$%
\Delta $ of the rest masses of the three quarks in a baryon (2$\Delta $ of
the quark and antiquark in a meson). Thus we can omit the binding energy -3$%
\Delta $ (or -2$\Delta $) and the corresponding rest mass parts 3$\Delta $
(or 2$\Delta $) of the quarks. This effect makes it appear as if there is no
strong binding energy in baryons (or mesons).

2). The energy band excited quarks u(313) and d(313) with $\overrightarrow{n}
$ = (0, 0, 0) will be short-lived quarks. They are, however, lowest energy
quarks. Since there is no lower energy position that they can decay into,
they are not short-lived quarks. Because they have the same rest mass and
intrinsic quantum numbers as the free excited quarks u(313) and d(313), they
cannot be distinguished from the free excited quarks by experiments. The
u(313) and d(313) with $\overrightarrow{n}$ = (0, 0, 0) will be covered up
by free excited u(313) and d(313) in experiments since the probability that
they are produced is much smaller than the probability that the free excited
u(313)-quark and d(313)-quark are produced. Therefore, we can omit u(313)
and d(313) with $\overrightarrow{n}$ = (0, 0, 0). There are only long-lived
and free excited u(313) and d(313) quarks in both theory and experiments. \
\ \ \ \ \ \ \ \ \ \ \ \ \ \ \ \ \ \ \ \ \ \ \ \ \ \ \ \ \ \ \ \ \ \ \ \ \ \
\ \ \ \ \ \ \ \ \ \ \ \ \ \ \ \ \ \ \ \ \ \ \ \ \ \ \ \ \ \ \ \ \ \ \ \ \ \
\ \ \ \ \ \ \ \ \ \ \ \ \ \ \ \ \ \ \ \ \ \ \ \ \ \ \ \ \ \ \ \ \ \ \ \ \ \
\ \ \ \ \ \ \ \ \ \ \ \ \ \ \ \ \ \ \ \ \ \ \ \ \ \ \ \ \ \ \ \ \ \ \ \ \ \
\ \ \ \ \ \ \ \ \ \ \ \ \ \ \ \ \ \ \ \ \ \ \ \ \ \ \ \ \ \ \ \ \ \ \ \ \ \
\ \ \ \ \ \ \ \ \ \ \ \ \ \ \ \ \ \ \ \ \ \ \ \ \ \ \ \ \ \ \ \ \ \ \ \ \ \
\ \ \ \ \ \ \ \ \ \ \ \ \ \ \ \ \ \ \ \ \ \ \ \ \ \ \ \ \ \ \ \ \ \ \ \ \ \
\ \ \ \ \ \ \ \ \ \ \ \ \ \ \ \ \ \ \ \ \ \ \ \ \ \ \ \ \ \ \ \ \ \ \ \ \ \
\ \ \ \ \ \ \ \ \ \ \ \ \ \ \ \ \ \ \ \ \ \ \ \ \ \ \ \ \ \ \ \ \ \ \ \ \ \
\ \ \ \ \ \ \ \ \ \ \ \ \ \ \ \ \ \ \ \ \ \ \ \ \ \ \ \ \ \ \ \ \ \ \ \ \ \
\ \ \ \ \ \ \ \ \ \ \ \ \ \ \ \ \ \ \ \ \ \ \ \ \ \ \ \ \ \ \ \ \ \ \ \ \ \
\ \ \ \ \ \ \ \ \ \ \ \ \ \ \ \ \ \ \ \ \ \ \ \ \ \ \ \ \ \ \ \ \ \ \ \ \ \
\ \ \ \ \ \ \ \ \ \ \ \ \ \ \ \ \ \ \ \ \ \ \ \ \ \ \ \ \ \ \ \ \ \ \ \ \ \
\ \ \ \ \ \ \ \ \ \ \ \ \ \ \ \ \ \ \ \ \ \ \ \ \ \ \ \ \ \ \ \ \ \ \ \ \ \
\ \ \ \ \ \ \ \ \ \ \ \ \ \ \ \ \ \ \ \ \ \ \ \ \ \ \ \ \ \ \ \ \ \ \ \ \ \
\ \ \ \ \ \ \ \ \ \ \ \ \ \ \ \ \ \ \ \ \ \ \ \ \ \ \ \ \ \ \ \ \ \ \ \ \ \
\ \ \ \ \ \ \ \ \ \ \ \ \ \ \ \ \ \ \ \ \ \ \ \ \ \ \ \ \ \ \ \ \ \ \ \ \ \
\ \ \ \ \ \ \ \ \ \ \ \ \ \ \ \ \ \ \ \ \ \ \ \ \ \ \ \ \ \ \ \ \ \ \ \ \ \
\ \ \ \ \ \ \ \ \ \ \ \ \ \ \ \ \ \ \ \ \ \ \ \ \ \ \ \ \ \ \ \ \ \ \ \ \ \
\ \ \ \ \ \ \ \ \ \ \ \ \ \ \ \ \ \ \ \ \ \ \ \ \ \ \ \ \ \ \ \ \ \ \ \ \ \
\ \ \ \ \ \ \ \ \ \ \ \ \ \ \ \ \ \ \ \ \ \ \ \ \ \ \ \ \ \ \ \ \ \ \ \ \ \
\ \ \ \ \ \ \ \ \ \ \ \ \ \ \ \ \ \ \ \ \ \ \ \ \ \ \ \ \ \ \ \ \ \ \ \ \ \
\ \ \ \ \ \ \ \ \ \ \ \ \ \ \ \ \ \ \ \ \ \ \ \ \ \ \ \ \ \ \ \ \ \ \ \ \ \
\ \ \ \ \ \ \ \ \ \ \ \ \ \ \ \ \ \ \ \ \ \ \ \ \ \ \ \ \ \ \ \ \ \ \ \ \ \
\ \ \ \ \ \ \ \ \ \ \ \ \ \ \ \ \ \ \ \ \ \ \ \ \ \ \ \ \ \ \ \ \ \ \ \ \ \
\ \ \ \ \ \ \ \ \ \ \ \ \ \ \ \ \ \ \ \ \ \ \ \ \ \ \ \ \ \ \ \ \ \ \ \ \ \
\ \ \ \ \ \ \ \ \ \ \ \ \ \ \ \ \ \ \ \ \ \ \ \ \ \ \ \ \ \ \ \ \ \ \ \ \ \
\ \ \ \ \ \ \ \ \ \ \ \ \ \ \ \ \ \ \ \ \ \ \ \ \ \ \ \ \ \ \ \ \ \ \ \ \ \
\ \ \ \ \ \ \ \ \ \ \ \ \ \ \ \ \ \ \ \ \ \ \ \ \ \ \ \ \ \ \ \ \ \ \ \ \ \
\ \ \ \ \ \ \ \ \ \ \ \ \ \ \ \ \ \ \ \ \ \ \ \ \ \ \ \ \ \ \ \ \ \ \ \ \ \
\ \ \ \ \ \ \ \ \ \ \ \ \ \ \ \ \ \ \ \ \ \ \ \ \ \ \ \ \ \ \ \ \ \ \ \ \ \
\ \ \ \ \ \ \ \ \ \ \ \ \ \ \ \ \ \ \ \ \ \ \ \ \ \ \ \ \ \ \ \ \ \ \ \ \ \
\ \ \ \ \ \ \ \ \ \ \ \ \ \ \ \ \ 

3). The five quarks of the current Quark Model correspond to the five
deduced ground quarks [u$\leftrightarrow $u(313), d$\leftrightarrow $d(313),
s$\leftrightarrow $d$_{s}$(493), c$\leftrightarrow $u$_{c}$(1753) and b$%
\leftrightarrow $d$_{b}$(4913)]. The current Quark Model uses only these
five quarks to explain baryons and mesons. In earlier times, this was
reasonable, natural and useful. Today, however, it is not reasonable since
physicists have discovered many high energy baryons and mesons that need
more high energy quarks to compose them.

\section{Conclusions\ \ \ \ \ \ \ \ \ \ \ \ \ \ \ \ \ \ \ \ \ \ \ \ \ \ \ \
\ \ \ \ \ \ }

1). There is only one elementary quark family $\epsilon $ with three colors
and two isospin states ($\epsilon _{u}$ with I$_{Z}$ = $\frac{1}{2}$ and Q =
+$\frac{2}{3}$, $\epsilon _{d}$ with I$_{Z}$ = $\frac{-1}{2}$ and Q = -$%
\frac{1}{3}$) for each color. Thus there are six Fermi (s = $\frac{1}{2}$)
elementary quarks with S = C =B = 0 in the vacuum. The elementary quarks $%
\epsilon _{u}$and $\epsilon _{d}$ have SU(2) symmetry.

2). All quarks in hadrons are the excited state of the elementary quark $%
\epsilon $. There are two types of excited states: free excited states and
energy band excited states. The free excited states are only the
u(313)-quark and the d(313)-quark. The energy band excited states are the
short-lived quarks, such as d$_{s}$(493), d$_{s}$(773), u$_{c}$(1753) and d$%
_{b}$(4913)....

3). There is a large binding energy -3$\Delta $ (or -2$\Delta $) among three
quarks in a baryon (or between the quark and the antiquark in a meson). It
may be a reason for the quark confinement.

4). Using the phenomenological formulae, we have deduced the rest masses and
intrinsic quantum numbers of quarks (Tables 3 and 4), baryons (Table 6) and
mesons (Table 7). The deduced intrinsic quantum numbers match the
experimental results \cite{Baryon04} and \cite{Meson04} exactly. The deduced
rest masses of the baryons are more than 98.5\% consistent with experimental
results \cite{Baryon04} and the deduced rest masses of the mesons are more
than 97\% consistent with experimental results \cite{Meson04}.

5). The current Quark Model is the five ground quark approximation of an
unborn, more fundamental model.

6). This paper predict some new quarks [d$_{S}$(773), d$_{S}$(1933) and $%
\text{u}_{C}$(6073)], baryons [$\Lambda _{C}$(6599) and $\Lambda _{b}$%
(9959)] and mesons [D(6231) and B(9502)].

\begin{center}
\bigskip \textbf{Acknowledgments}
\end{center}

I sincerely thank Professor Robert L. Anderson for his valuable advice. I
acknowledge\textbf{\ }my indebtedness to Professor D. P. Landau for his help
also. I would like to express my heartfelt gratitude to Dr. Xin Yu for
checking the calculations. I sincerely thank Professor Yong-Shi Wu for his
important advice and help. I thank Professor Wei-Kun Ge for his support and
help. I sincerely thank Professor Kang-Jie Shi for his advice.

\bigskip\ \ \ \ \ \ \ \ \ \ \ \ \ \ \ \ \ \ \ 


\begin{thebibliography}{9}
\bibitem{Planck} R. M. Eisberg, Fundamentals of Mondern Physics, John Wiley
\& Sons, Inc, 64 (1961).\textit{\ }

\bibitem{Standard} M. K. Gaillard, P. D. Grannis, and F. J \ Sciulli, Rev.
Mod. Phys., \textbf{71} No. 2 Centenary, S96 (1999).

\bibitem{Quark Model} M. Gell-Mann, Phys. Lett.\ \textbf{8,} 214 (1964); \
G. Zweig, CERN Preprint CERN-Th-401, CERN-Th-412 (1964); Particle Data
Group, Phys. Lett. \textbf{B592}, 154 (2004)

\bibitem{0502091} J. L. Xu, hep-ph/0502091.

\bibitem{Quarks} Particle Data Group, Phys. Lett. \textbf{B592}, 37 (2004).

\bibitem{Baryon04} Particle Data Group, Phys. Lett. \textbf{B592}, 66--78
(2004).

\bibitem{Meson04} Particle Data Group, Phys. Lett. \textbf{B592}, 38--65
(2004). \ \ \ 
\end{thebibliography}
\end{document}